\pdfoutput=1

\documentclass[11pt,twoside,a4paper,cmspaper,final,collab]{cms-tdr}
\def\svnVersion{6ebe590-D}\def\svnDate{2019/05/03}\def\cmsCernNoTag{CERN-EP-2018-353}\def\cmsCernDate{\today}\def\cmsMessage{"Published in the European Physical Journal C as \href{http://dx.doi.org/10.1140/epjc/s10052-019-6926-x}{\doi{10.1140/epjc/s10052-019-6926-x}.}"}

\begin{document}\cmsNoteHeader{SUS-18-002}

\hyphenation{had-ron-i-za-tion}
\hyphenation{cal-or-i-me-ter}
\hyphenation{de-vices}
\RCS$HeadURL$
\RCS$Id$

\newlength\cmsFigWidth
\ifthenelse{\boolean{cms@external}}{\setlength\cmsFigWidth{0.35\textwidth}}{\setlength\cmsFigWidth{0.45\textwidth}}
\ifthenelse{\boolean{cms@external}}{\providecommand{\cmsLeft}{upper\xspace}}{\providecommand{\cmsLeft}{left\xspace}}
\ifthenelse{\boolean{cms@external}}{\providecommand{\cmsRight}{lower\xspace}}{\providecommand{\cmsRight}{right\xspace}}
\newlength\cmsTabSkip\setlength\cmsTabSkip{-1.0ex}
\ifthenelse{\boolean{cms@external}}{\providecommand{\cmsTable}[1]{#1}}{\providecommand{\cmsTable}[1]{\resizebox{\textwidth}{!}{#1}}}

\newcommand{\nj}{\ensuremath{N_{\text{jets}}}\xspace}
\newcommand{\nb}{\ensuremath{N_{{\cPqb}\text{-jets}}}\xspace}
\newcommand{\htg}{\ensuremath{\HT^{\cPgg}}\xspace}
\newcommand{\ptg}{\ensuremath{\pt^{\cPgg}}\xspace}
\newcommand{\ptj}{\ensuremath{\pt^{\text{jet}}}\xspace}
\newcommand{\ptl}{\ensuremath{\pt^{\ell}}\xspace}
\newcommand{\pttk}{\ensuremath{\pt^{\text{track}}}\xspace}
\newcommand{\vecpttk}{\ensuremath{\ptvec^{\kern1pt\text{track}}}\xspace}
\newcommand{\qmult}{\ensuremath{Q_{\text{mult}}}\xspace}
\newcommand{\rhl}{\ensuremath{R_{\text{h/l}}}\xspace}
\newcommand{\dphi}{\ensuremath{\Delta\phi}\xspace}
\newcommand{\deta}{\ensuremath{\Delta\eta}\xspace}
\newcommand{\deltar}{\ensuremath{\Delta R}\xspace}
\newcommand{\tauhad}{\ensuremath{\tauh}\xspace}
\newcommand{\zg}{\ensuremath{\cPZ{\cPgg}}\xspace}
\newcommand{\znn}{\ensuremath{\cPZ(\cPgn\cPagn)}\xspace}
\newcommand{\znng}{\ensuremath{\cPZ(\cPgn\cPagn)\cPgg}\xspace}
\newcommand{\zllg}{\ensuremath{\cPZ(\ell^{+}\ell^{-})\cPgg}\xspace}
\newcommand{\llg}{\ensuremath{\ell^{+}\ell^{-}\cPgg}\xspace}
\newcommand{\eleg}{\ensuremath{\Pe{\cPgg}}\xspace}
\newcommand{\mug}{\ensuremath{\Pgm{\cPgg}}\xspace}
\newcommand{\Wg}{\ensuremath{\PW{\cPgg}}\xspace}
\newcommand{\ttg}{\ensuremath{\ttbar{\cPgg}}\xspace}
\newcommand{\tg}{\ensuremath{{\cPqt}{\cPgg}}\xspace}
\newcommand{\gjets}{\ensuremath{{\cPgg}\text{+jets}}\xspace}
\newcommand{\Wgjets}{{\PW}\gjets}
\newcommand{\PV}{\ensuremath{\cmsSymbolFace{V}}\xspace}
\newcommand{\Vgjets}{{\PV}\gjets}
\newcommand{\ttjets}{{\ttbar}+jets\xspace}
\newcommand{\Wjets}{{\PW}+jets\xspace}
\newcommand{\Zjets}{{\PZ}+jets\xspace}
\newcommand{\Rnnll}{\ensuremath{R_{\PGn\PGn/\ell\ell}}\xspace}
\newcommand{\mnlsp}{\ensuremath{m_{\PSGczDo}}\xspace}
\newcommand{\mgluino}{\ensuremath{m_{\PSg}}\xspace}
\newcommand{\mstop}{\ensuremath{m_{\PSQt}}\xspace}
\newcommand{\mgravitino}{\ensuremath{m_{\PXXSG}}\xspace}
\newcommand{\mdilep}{\ensuremath{m_{\ell\ell}}\xspace}
\newcommand{\mz}{\ensuremath{m_{\PZ}}\xspace}
\newcommand{\njb}{\ensuremath{N_{\mathrm{j}}^{\mathrm{b}}}\xspace}
\newcommand{\deltam}{\ensuremath{\Delta m}\xspace}
\newcommand{\Itrack}{\ensuremath{I_{\text{track}}}\xspace}
\newcommand{\Msb}{\ensuremath{M_{\text{SB}}}\xspace}
\newcommand{\Mplanck}{\ensuremath{M_{\text{Pl}}}\xspace}
\newcommand{\Riso}{\ensuremath{R_{I}}\xspace}
\newcommand{\TFmt}{\ensuremath{T_{\Pgm,\Pgt}}\xspace}
\newcommand{\TFe}{\ensuremath{T_{\Pe}}\xspace}
\newcommand{\muR}{\ensuremath{\mu_{\text{R}}}\xspace}
\newcommand{\muF}{\ensuremath{\mu_{\text{F}}}\xspace}
\newcommand{\abseta}{\ensuremath{\abs{\eta}}}

\cmsNoteHeader{SUS-18-002}
\title{Search for supersymmetry in events with a photon, jets, {\cPqb}-jets, and missing transverse momentum in proton-proton collisions at 13\TeV}
\titlerunning{Search for supersymmetry in events with a photon, jets, {\cPqb}-jets, and missing transverse momentum at 13\TeV}

\date{\today}

\abstract{A search for supersymmetry is presented based on events with at least
one photon, jets, and large missing transverse momentum produced in proton-proton collisions
at a center-of-mass energy of 13\TeV. The data correspond to an integrated
luminosity of 35.9\fbinv and were recorded at the LHC with the CMS detector in 2016.
The analysis characterizes signal-like events by categorizing the data into various
signal regions based on the number of jets, the number of {\cPqb}-tagged jets, and the missing
transverse momentum. No significant excess of events is observed with respect to the expectations
from standard model processes. Limits are placed on the gluino and top squark pair production cross sections
using several simplified models of supersymmetric
particle production with gauge-mediated supersymmetry breaking. Depending on the model and the
mass of the next-to-lightest supersymmetric particle, the production of gluinos with masses as large as 2120\GeV
and the production of top squarks with masses as large as 1230\GeV are excluded at 95\% confidence level.
}

\hypersetup{%
pdfauthor={CMS Collaboration},%
pdftitle={Search for supersymmetry in events with a photon, jets, b-jets, and missing transverse momentum in proton-proton collisions at 13 TeV},%
pdfsubject={CMS},%
pdfkeywords={CMS, physics, photons, supersymmetry, b tagging, GMSB}}

\maketitle

\section{Introduction}
\label{sec:introduction}

The standard model (SM) of particle physics successfully describes many
phenomena, but lacks several necessary elements to provide a complete description of nature, including a source for the relic abundance
of dark matter (DM)~\cite{Zwicky:1933gu,Rubin:1970zza} in the universe. In addition, the SM must resort to fine tuning~\cite{Barbieri:1987fn,Dimopoulos:1995mi,Barbieri:2009ev,Papucci:2011wy}
to explain the hierarchy between the Planck mass scale and the electroweak scale set by the vacuum expectation value of the Higgs field, the existence of which was recently confirmed by the observation of the
Higgs boson (\PH)~\cite{Sirunyan:2017exp,Aad:2015zhl}.
Supersymmetry (SUSY)~\cite{Ramond:1971gb,Golfand:1971iw,Neveu:1971rx,
Volkov:1972jx,Wess:1973kz,Wess:1974tw,Fayet:1974pd,Nilles:1983ge} is an extension of the SM that can provide both a viable DM candidate
and additional particles that inherently cancel large
quantum corrections to the Higgs boson mass-squared term from the SM fields.

Supersymmetric models predict a bosonic superpartner for each SM fermion and a fermionic superpartner for each SM boson;
each new particle's spin differs from that of its SM partner by half a unit.
SUSY also includes a second Higgs doublet.
New colored states, such as gluinos ({\PSg}) and top squarks
({\PSQt}), the superpartners of the gluon and the top
quark, respectively, are expected
to have masses on the order of 1\TeV to avoid fine tuning in the SM Higgs boson mass-squared term.
In models that conserve $R$-parity~\cite{bib-rparity},
each superpartner carries a conserved quantum
number that requires superpartners to be produced
in pairs and causes the lightest SUSY particle (LSP) to
be stable. The stable LSP can serve as a DM candidate.

The signatures targeted in this paper are motivated by models in which gauge-mediated SUSY breaking (GMSB)
is responsible for separating the masses of the SUSY particles from those of their SM counterparts.
In GMSB models, the gaugino masses are expected to be proportional to the size of their fundamental couplings.
This includes the superpartner of the graviton, the gravitino (\PXXSG), whose mass is proportional to $\Msb/\Mplanck$,
where $\Msb$ represents the scale of the SUSY breaking interactions and $\Mplanck$ is the Planck scale where gravity is expected to become strong.
GMSB permits a significantly lower symmetry-breaking scale than, e.g., gravity mediation,
and therefore generically predicts that the \PXXSG is the LSP~\cite{Meade:2008wd,PhysRevLett.38.1433,CREMMER1978231}, with a mass often much less than 1\GeV.
Correspondingly, the next-to-LSP (NLSP) is typically a neutralino,
a superposition of the superpartners of the neutral bosons. The details of the quantum numbers of the NLSP
play a large part in determining the phenomenology of GMSB models, including the relative frequencies of the Higgs bosons, {\PZ} bosons, and photons
produced in the NLSP decay.

\begin{figure*}[htb]
\centering
\hspace*{\fill}{\includegraphics[width=\cmsFigWidth]{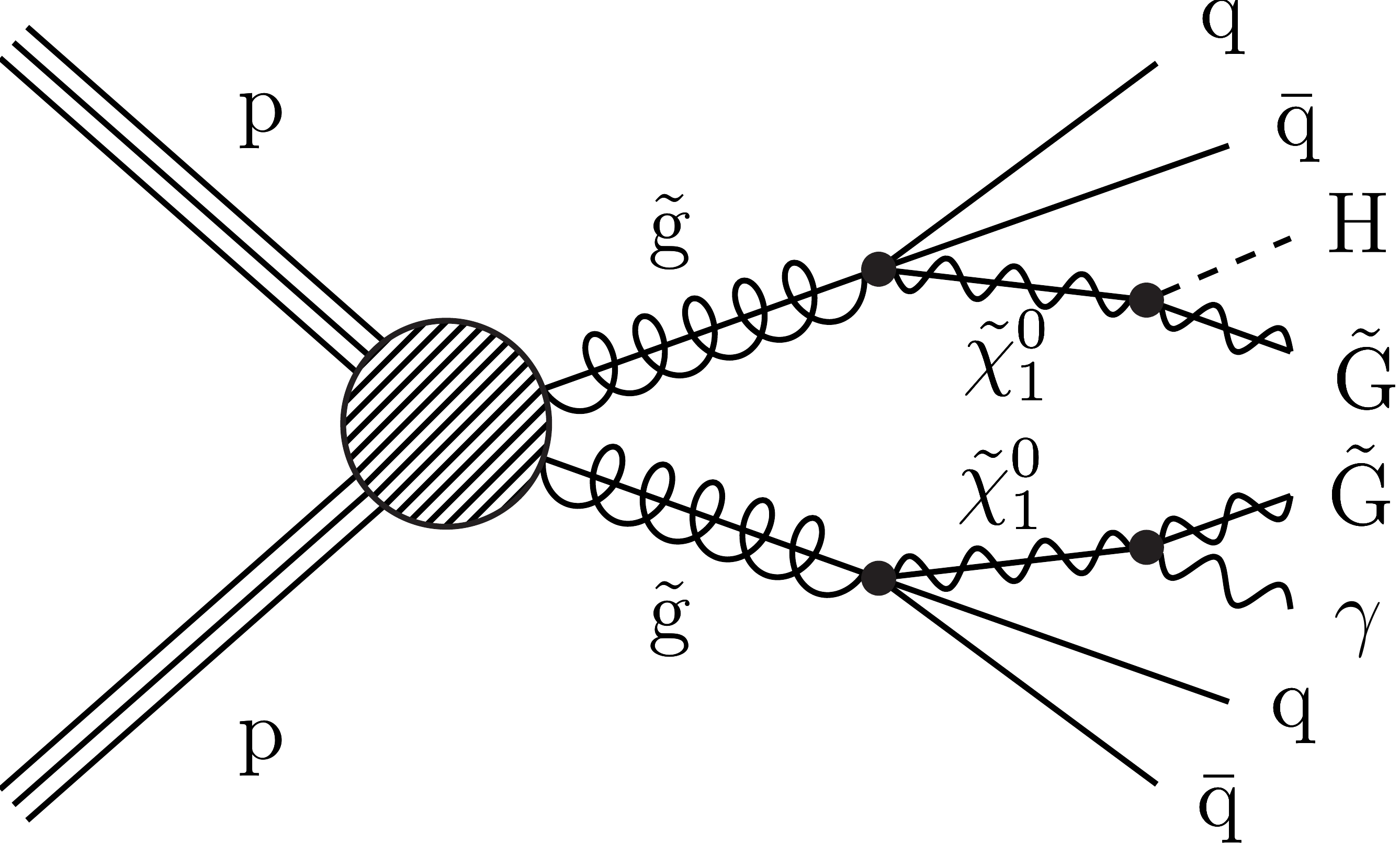}}\hfill{\includegraphics[width=\cmsFigWidth]{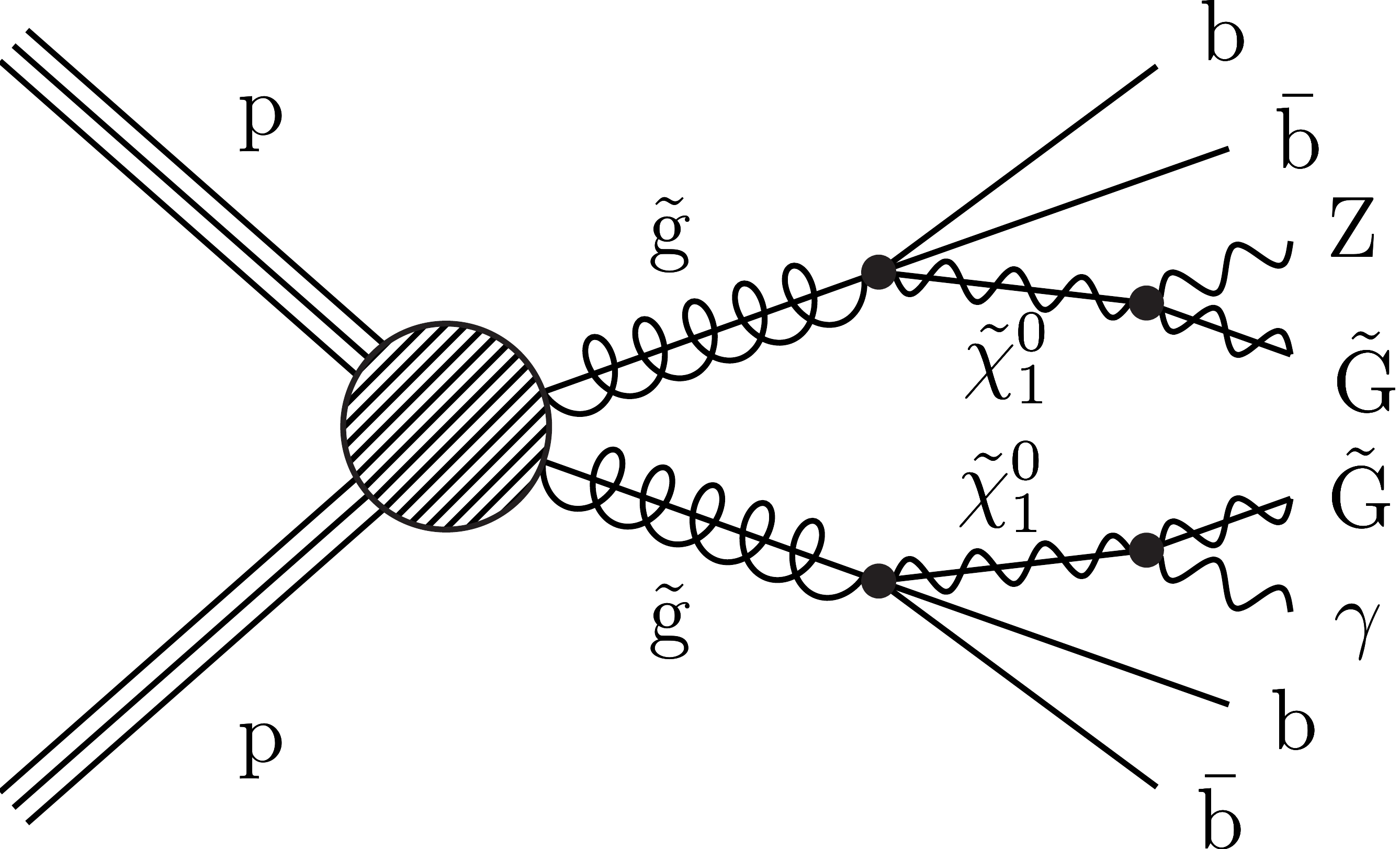}}\hspace*{\fill}\null\\
\vspace{0.5cm}
\hspace*{\fill}{\includegraphics[width=\cmsFigWidth]{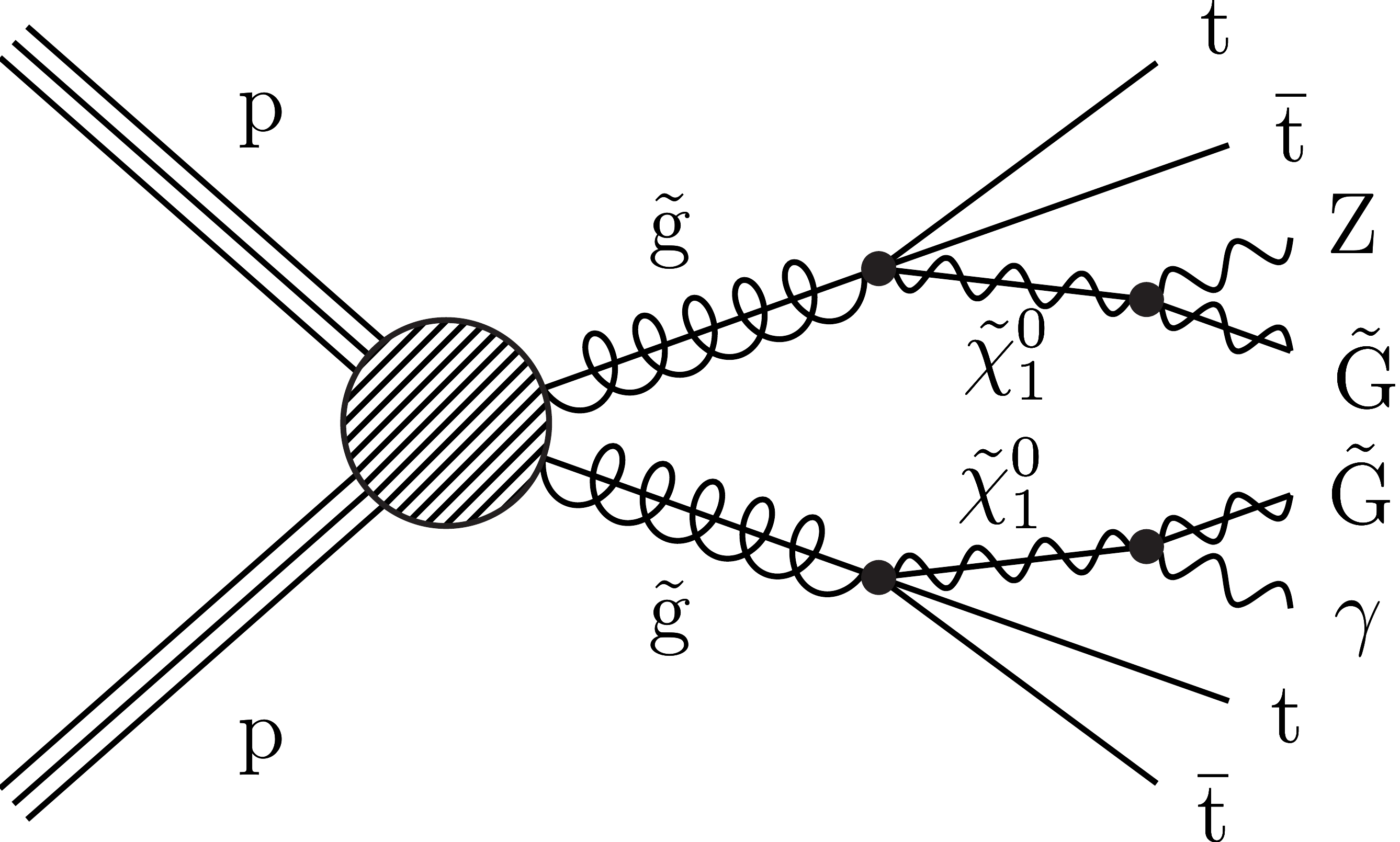}}\hfill{\includegraphics[width=\cmsFigWidth]{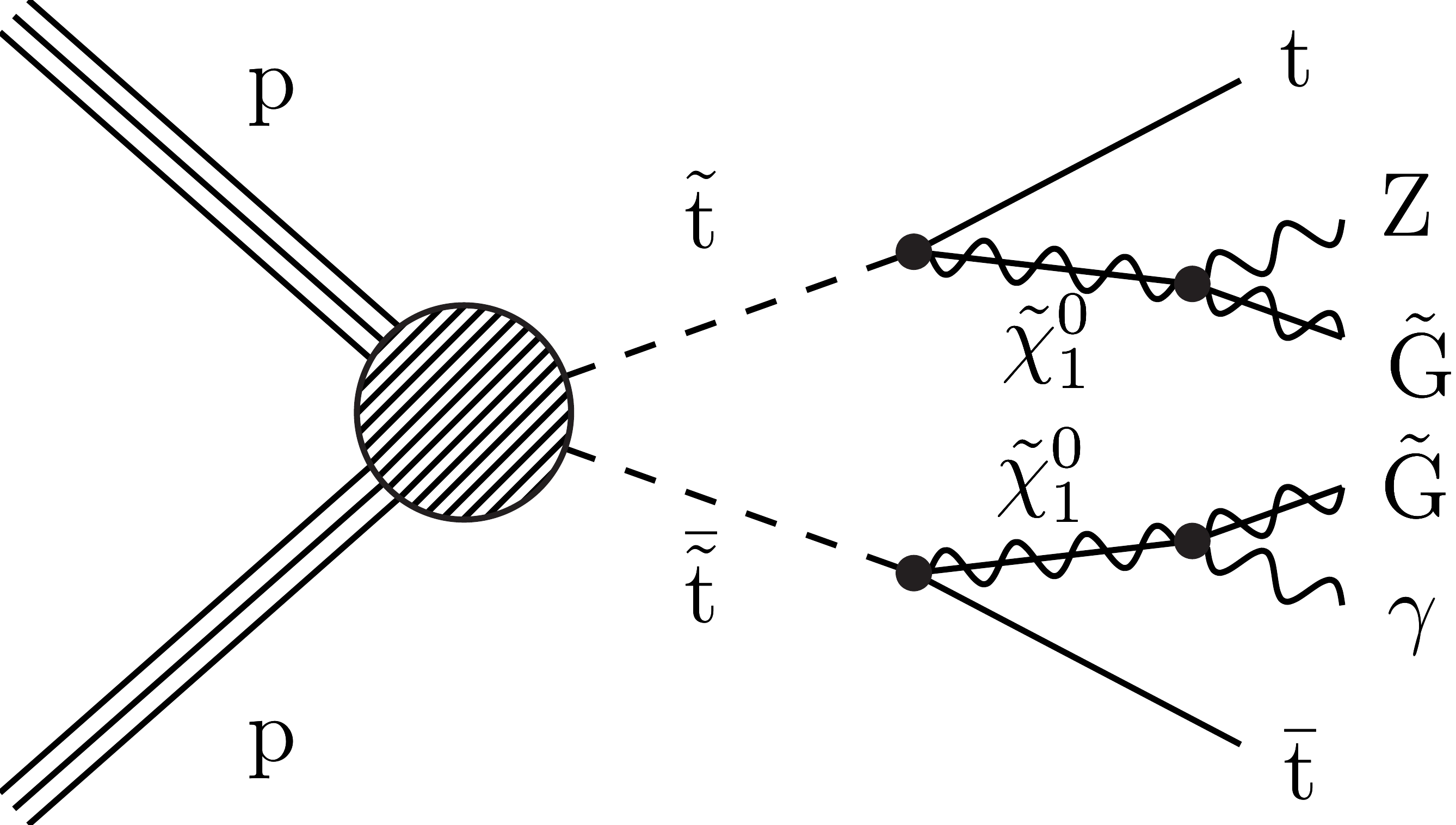}}\hspace*{\fill}
\caption{Example diagrams depicting the simplified models
used, which are defined in the text. The top left diagram depicts the T5qqqqHG model, the
top right diagram depicts the T5bbbbZG model, the bottom left
diagram depicts the T5ttttZG model, and the bottom right
depicts the T6ttZG model.}
\label{fig:SMS_diagram}
\end{figure*}

The scenario of a natural SUSY spectrum with GMSB and $R$-parity conservation typically manifests
as events with multiple jets, at least one photon, and large \ptmiss, the magnitude of the missing transverse momentum.
Depending on the topology, these jets can arise from either light-flavored quarks (\cPqu, \cPqd, \cPqs, \cPqc) or {\cPqb}
quarks. We study four simplified models~\cite{bib-sms-1,bib-sms-2,bib-sms-3,bib-sms-4,Chatrchyan:2013sza};
example diagrams depicting these models are shown in Fig.~\ref{fig:SMS_diagram}.
Three models involve gluino pair production (prefixed with T5), and one model involves top squark pair production (prefixed with T6).
In the T5qqqqHG model, each gluino decays to a pair of light-flavored quarks (\qqbar) and a
neutralino (\PSGczDo).  The T5bbbbZG and T5ttttZG models are similar to T5qqqqHG, except that the
each pair of light-flavored quarks is replaced by a pair of bottom quarks (\bbbar) or a pair of top quarks
(\ttbar), respectively. In the T5qqqqHG model, the \PSGczDo decays either to an SM Higgs boson and a \PXXSG or to a photon and a \PXXSG.  The
$\PSGczDo\to\PH\PXXSG$ branching fraction is assumed to be 50\%, and the smallest \PSGczDo mass considered is 127\GeV.
In the T5bbbbZG and T5ttttZG models, the neutralinos decay to
$\PZ\PXXSG$ and $\cPgg\PXXSG$ with equal probability.
The T6ttZG model considers top squark pair production, with each top squark decaying into a top quark and a neutralino.
The neutralino can then decay with equal probability to a photon and a \PXXSG or to a {\PZ} boson and a \PXXSG.
For the models involving the decay $\PSGczDo\to\PZ\PXXSG$, we probe \PSGczDo masses down to 10\GeV.
All decays of SUSY particles are assumed to be prompt.
In all models, the mass \mgravitino is fixed to be 1\gev, to be consistent with other published results.
For the parameter space explored here, the kinematic properties do not depend strongly on the exact value of \mgravitino.

\begin{sloppypar} The proton-proton ($\Pp\Pp$) collision data used in this search correspond to an integrated luminosity of 35.9\fbinv and were collected with the
CMS detector during the 2016 run of the CERN LHC~\cite{CMS-PAS-LUM-17-001}.
Signal-like events with at least one photon are classified into signal regions depending on
the number of jets \nj, the number of tagged
bottom quark jets \nb, and the \ptmiss.  The expected
yields from SM backgrounds are estimated using a combination of simulation
and data control regions. We search for gluino or top squark pair production
as an excess of observed data events compared to the expected background yields.
\end{sloppypar}

Previous searches for $R$-parity conserving SUSY with photons in the final state performed
by the CMS Collaboration are documented in Refs.~\cite{Sirunyan:2017nyt,Sirunyan:2017yse}.
Similar searches have also been performed by the ATLAS
Collaboration~\cite{Aad:2015hea,ATLASCollaboration:2016wlb,ATLASCollaboration:2018ggm}.
This work improves on the previous results by identifying jets from {\cPqb} quarks, which can be produced by all of the signal models shown in Fig.~\ref{fig:SMS_diagram}.
We also include additional signal regions that
exploit high jet multiplicities for sensitivity to
high-mass gluino models, and we rely more on observed data for the background
estimations.
These improvements enable us to explore targeted signal models that produce {\cPqb} quarks in the final state
and are expected to improve sensitivity to the models explored in Refs.~\cite{Sirunyan:2017nyt,Sirunyan:2017yse,Aad:2015hea,ATLASCollaboration:2016wlb,ATLASCollaboration:2018ggm}.

In this paper, a description of the CMS detector and simulation used are presented in Section~\ref{sec:detector}.
The event reconstruction and signal region selections are presented in Section~\ref{sec:event_reconstruction_selection}.
The methods used for predicting the
SM backgrounds are presented in Section~\ref{sec:sm_backgrounds}. Results are
given in Section~\ref{sec:results}.  The analysis is summarized in
Section~\ref{sec:summary}.

\section{Detector and simulation}
\label{sec:detector}

\begin{sloppypar}A detailed description of the CMS detector, along with a definition of the coordinate system
and pertinent kinematic variables, is given in Ref.~\cite{Chatrchyan:2008aa}. Briefly, a cylindrical superconducting
solenoid with an inner diameter of 6\unit{m} provides a 3.8\unit{T} axial magnetic field. Within the
cylindrical volume are a silicon pixel and strip tracker, a lead tungstate crystal electromagnetic
calorimeter (ECAL), and a brass and scintillator hadron calorimeter (HCAL). The tracking
detectors cover the pseudorapidity range ${\abseta<2.5}$. The ECAL and HCAL, each composed of a
barrel and two endcap sections, cover ${\abseta<3.0}$. Forward calorimeters extend the coverage to
${3.0<\abseta<5.0}$. Muons are detected within ${\abseta<2.4}$ by gas-ionization detectors embedded
in the steel flux-return yoke outside the solenoid. The detector is nearly hermetic, permitting
accurate measurements of \ptmiss.  The CMS trigger is described in Ref.~\cite{Khachatryan:2016bia}.
\end{sloppypar}

Monte Carlo (MC) simulation is used to design the analysis, to provide input for background
estimation methods that use data control regions, and to predict event rates from simplified models.
Simulated SM background processes include jets produced through the strong interaction, referred to as quantum chromodynamics (QCD) multijets, \ttjets, \Wjets, \Zjets, \gjets, {\ttg}, {\tg}, and \Vgjets ({\PV} = \cPZ, {\PW}).
The SM background events are generated using the \MGvATNLO v2.2.2 or v2.3.3 generator~\cite{Alwall:2014hca,Kalogeropoulos:2018cke,Artoisenet:2012st} at leading order (LO) in perturbative QCD,
except {\ttg} and {\tg}, which are generated at next-to-leading order (NLO).
The cross sections used for normalization are computed at NLO or next-to-NLO~\cite{Alwall:2014hca,Czakon:2011xx,Gavin:2012sy,Gavin:2010az}.
The QCD multijets, diboson ({\PV}\cPgg), top quark, and vector boson plus jets events are generated with
up to two, two, three, and four additional partons in the matrix element calculations, respectively.
Any duplication of events between pairs of related processes---QCD multijets and \gjets; \ttjets and {\ttg}; \Wjets and \Wgjets---is removed using generator information.

\begin{sloppypar} The NNPDF3.0~\cite{Ball:2014uwa} LO (NLO) parton distribution functions (PDFs) are used for
samples simulated at LO (NLO).  Parton showering and hadronization are described using the \PYTHIA 8.212
generator~\cite{Sjostrand:2014zea} with the CUETP8M1 underlying event
tune~\cite{Khachatryan:2015pea}.  Partons generated with \MGvATNLO
and \PYTHIA that would otherwise be counted twice are removed using the MLM~\cite{Alwall:2007fs}
and \textsc{FxFx}~\cite{Frederix:2012ps} matching schemes in LO and NLO samples, respectively.
\end{sloppypar}

Signal samples are simulated at LO using the
\MGvATNLO v2.3.3 generator and their yields are normalized using NLO plus next-to-leading logarithmic (NLL) cross sections~\cite{bib-nlo-nll-01,bib-nlo-nll-02,bib-nlo-nll-03,bib-nlo-nll-04,bib-nlo-nll-05}.
The decays of gluinos, top squarks, and neutralinos are modeled with \PYTHIA.

The detector response to particles produced in the simulated collisions is modeled with the
\GEANTfour~\cite{Agostinelli:2002hh} detector simulation package for SM processes. Because of
the large number of SUSY signals considered, with various gluino, squark, and neutralino masses,
the detector response for these processes is simulated with the CMS fast simulation~\cite{Abdullin:2011zz,Giammanco:2014bza}.
The results from the fast simulation generally agree with the results from the full simulation. Where there is disagreement,
corrections are applied, most notably a correction of up to 10\% to adjust for differences in the modeling of \ptmiss.

\section{Event reconstruction and selection}
\label{sec:event_reconstruction_selection}

The CMS particle-flow (PF) algorithm~\cite{CMS-PRF-14-001} aims to reconstruct every particle in each event,
using an optimal combination of information from all detector systems. Particle candidates are identified as
charged hadrons, neutral hadrons, electrons, photons, or muons.
For electron and photon PF candidates, further requirements are applied to the ECAL shower shape and the ratio of associated energies in the ECAL and HCAL~\cite{Khachatryan:2015hwa,CMS:EGM-14-001}.
Similarly, for muon PF candidates, further requirements are applied to the matching between track segments in the silicon tracker and the muon detectors~\cite{Sirunyan:2018fpa}.
These further requirements improve the quality of the reconstruction. Electron and muon candidates are restricted to
$\abseta<2.5$ and $<2.4$, respectively. The \ptmiss is calculated as the magnitude of the negative vector \pt sum of all PF candidates.

After all interaction vertices are reconstructed, the primary $\Pp\Pp$ interaction vertex is selected
as the vertex with the largest $\pt^2$ sum of all physics objects.
The physics objects used in this calculation are produced by a jet-finding
algorithm~\cite{Cacciari:2008gp,Cacciari:2011ma} applied to all charged-particle tracks associated to the vertex, plus the corresponding
\ptmiss computed from those jets.
To mitigate the effect of secondary $\Pp\Pp$ interactions (pileup), charged-particle tracks associated
with vertices other than the primary vertex are not considered for jet clustering or calculating object isolation sums.

Jets are reconstructed by clustering PF candidates using the anti-\kt jet algorithm~\cite{Cacciari:2008gp,Cacciari:2011ma} with a size
parameter of 0.4. To eliminate spurious jets, for example those induced by electronics noise, further jet quality criteria~\cite{CMS-PAS-JME-16-003} are applied.
The jet energy response is corrected for the nonlinear response of the detector~\cite{Khachatryan:2016kdb}.
There is also a correction to account for the expected contributions of neutral particles from pileup, which cannot be removed based on association with secondary vertices~\cite{Cacciari:2007fd}.
Jets are required to have $\pt>30\gev$ and are restricted to be within $\abseta < 2.4$. The combined secondary vertex algorithm (CSVv2) at the medium working point~\cite{BTV-16-002}
is applied to each jet to determine if it should be identified as a bottom quark jet. The CSVv2 algorithm at the specified working point has a 55\% efficiency to correctly identify
{\cPqb} jets with $\pt \approx 30\gev$. The corresponding misidentification probabilities are 1.6\% for gluon and light-flavor quark jets, and 12\% for charm quark jets.

Photons with $\pt>100\GeV$ and $\abseta<2.4$ are used in this analysis, excluding the ECAL transition region with ${1.44 < \abseta < 1.56}$.
To suppress jets erroneously identified as photons from neutral hadron decays, photon candidates
are required to be isolated. An isolation cone of radius $\deltar = \sqrt{\smash[b]{(\dphi)^2 + (\deta)^2}} < 0.2$ is used, with no dependence on
the \pt of the photon candidate. Here, $\phi$ is the azimuthal angle in radians. The energy measured in the isolation cone
is corrected for contributions from pileup~\cite{Cacciari:2007fd}.  The shower shape and the fractions of hadronic and electromagnetic
energy associated with the photon candidate are required to be consistent with expectations
from prompt photons. The candidates matched to a track measured by the pixel detector
(pixel seed) are rejected because they are likely to result from electrons that produced electromagnetic showers.

Similarly, to suppress jets erroneously identified as leptons and genuine leptons from hadron decays,
electron and muon candidates are also subjected to isolation requirements.
The isolation variable $I$ is computed from the scalar \pt sum of selected charged hadron, neutral hadron, and photon PF candidates,
divided by the lepton \pt. PF candidates enter the isolation sum if they satisfy $R < \Riso(\pt)$.
The cone radius $\Riso$ decreases with lepton \pt because the collimation of the decay products of the parent particle of the lepton
increases with the Lorentz boost of the parent~\cite{Rehermann:2010vq}. The values used are $\Riso = 0.2$ for $\ptl<50\GeV$, $\Riso = 10\GeV/\ptl$ for $50\leq\ptl\leq200\GeV$,
and $\Riso = 0.05$ for $\ptl>200\GeV$, where $\ell=\Pe,\Pgm$.
As with photons, the expected contributions from pileup are subtracted from the isolation variable.
The isolation requirement is ${I<0.1\,(0.2)}$ for electrons (muons).

We additionally veto events if they contain PF candidates which are identified as an electron, a muon, or
a charged hadron, and satisfy an isolation requirement computed using tracks.
Isolated hadronic tracks are common in background events with a tau lepton that decays hadronically.
The track isolation variable \Itrack is computed for each candidate from the scalar \pt sum of selected other charged-particle tracks,
divided by the candidate \pt. Other charged-particle tracks are selected if they lie within a cone of radius 0.3 around the candidate direction and come from the primary vertex.
The isolation variable must satisfy $\Itrack < 0.2$ for electrons and muons, and $\Itrack < 0.1$ for charged hadrons.
Isolated tracks are required to satisfy $\abseta<2.4$, and the transverse mass of each isolated track with \ptmiss, $\mT=\sqrt{\smash[b]{2\pttk\ptmiss(1-\cos{\dphi})}}$
where $\dphi$ is the difference in $\phi$ between \vecpttk and \ptvecmiss, is required to be less than 100\gev.

Signal event candidates were
recorded by requiring a photon at the trigger level with a requirement
${\ptg>90\gev}$ if ${\htg=\ptg+\Sigma \ptj>600\gev}$ and ${\ptg>165\gev}$ otherwise.
These quantities are computed at the trigger level.
The efficiency of this trigger, as measured in data, is $(98\pm2)\%$ after applying the selection criteria described below.
Additional triggers, requiring the
presence of charged leptons, photons, or minimum $\HT=\Sigma \ptj$, are used to
select control samples employed in the evaluation of backgrounds.

\begin{sloppypar} Signal-like candidate events must fulfill one of two requirements, based on the trigger criteria described above:
${\ptg>100\gev}$ and ${\htg>800\gev}$, or $\ptg>190\gev$ and $\htg>500\gev$.
In addition to these requirements, the events should have at least 2 jets and $\ptmiss>100\gev$.
To reduce backgrounds from the SM processes
that produce a leptonically decaying {\PW} boson, resulting in \ptmiss from the undetected neutrino, events are rejected if they
have any charged light leptons (\Pe, \Pgm) with $\pt>10\gev$ or any isolated electron, muon, or charged hadron tracks with ${\pt>5,5,10\gev}$, respectively.
Events from the \gjets process typically satisfy the above criteria when the energy of a jet is mismeasured, inducing artificial \ptmiss.
To reject these events, the two highest \pt jets
are both required to have an angular separation from the \ptmiss direction in the transverse plane, $\dphi_{1,2}>0.3$.
Events with reconstruction failures, detector noise, or beam halo interactions are rejected using dedicated identification
requirements~\cite{CMS-PAS-JME-16-004}.
\end{sloppypar}

\begin{sloppypar} The selected events are divided into 25 exclusive signal regions, also called signal bins, based on
\ptmiss, the number of jets \nj, and the number of {\cPqb}-tagged jets \nb.  The signal regions
can be grouped into 6 categories based on \nj and \nb, whose intervals are defined to be
\nj: 2--4, 5--6, ${\geq}7$; and \nb: 0, ${\geq}1$.  Within
each of the 6 categories, events are further distinguished based on 4 exclusive
regions, defined as: ${200<\ptmiss<270}$, ${270<\ptmiss<350}$, ${350<\ptmiss<450}$, and
${\ptmiss>450\gev}$.  In the lowest \nj, \nb category, the highest \ptmiss bin is further subdivided
into two intervals: ${450<\ptmiss<750}$ and ${\ptmiss>750\gev}$. Events with ${100<\ptmiss<200\gev}$
are used as a control region for estimating SM backgrounds. These categories in \nj, \nb, and \ptmiss
were found to provide good sensitivity to the various signal models described above, while minimizing
uncertainties in the background predictions.
\end{sloppypar}

\section{Background estimation}
\label{sec:sm_backgrounds}

There are four main mechanisms by which SM processes can produce events with the target
signature of a photon, multiple jets, and \ptmiss.  These mechanisms are: (1) the production of
a high-\pt photon along with a {\PW} or {\PZ} boson that decays leptonically, and either any resulting electron
or muon is ``lost'' (lost-lepton) or any resulting $\Pgt$ lepton decays hadronically (\tauh); (2) the production of
a {\PW} boson that decays to $\Pe\nu$ and the electron is misidentified as a photon; (3) the production of
a high-\pt photon in association with a {\PZ} boson that decays to neutrinos; and (4) the production of a photon
along with a jet that is mismeasured, inducing high \ptmiss. QCD multijet events with a jet misidentified as a photon
and a mismeasured jet do not contribute significantly to the SM background.

The total event yield from each source of background is estimated separately for each of the 25 signal regions.
The methods and uncertainties associated with the background predictions are detailed
in the following sections.

\subsection{Lost-lepton and \texorpdfstring{\tauhad}{hadronic tau} backgrounds}

The lost-lepton background arises from events in which the charged lepton from a leptonically
decaying {\PW} boson, produced directly or from the decay of a top quark, cannot be identified. This can occur because the lepton is out of acceptance, fails the identification requirements,
or fails the isolation requirements. For example, in events with high-\pt top quarks, the top quark decay
products will be collimated, forcing the {\cPqb} jet to be closer to the charged lepton. In this case,
the lepton is more likely to fail the isolation requirements. This background is estimated by studying
control regions in both data and simulation, obtained by requiring both a well-identified photon and a light lepton (\Pe, \Pgm). For every signal region, there are two lost lepton control regions that have the exact same definition as the signal region except either exactly one electron or exactly one muon is required.

The \tauhad background arises from events in which a {\PW} boson decays to a {\Pgt} lepton, which subsequently decays to
mesons and a neutrino. These hadronic decays of {\Pgt} leptons occur approximately $65\%$ of the time. Because of lepton universality, the fraction
of events with \tauhad candidates can be estimated from the yield of events containing a single muon, after correcting for
the reconstruction differences and for the \tauhad branching fraction.

The lost-lepton and \tauhad background predictions rely on an extrapolation between \eleg or \mug event yields and single
photon event yields.  In all control regions where a single light lepton is required, the dominant SM processes that contribute are \Wg
and \ttg.  Lost-muon and hadronic tau events are estimated using
\mug control regions, while lost-electron events are estimated using \eleg control regions.  In each control region, exactly one electron
or muon is required and the isolated track veto for the selected lepton flavor is removed.
In order to reduce the effect of signal contamination and to increase the fraction of SM events in the control sample, events are only selected if
the \mT of the lepton-\ptmiss system
is less than 100\gev. In SM background events with a single lepton and \ptmiss, the \mT
of the system is constrained by the mass of the {\PW} boson; this is not the case for signal events,
because of the presence of gravitinos.
All other kinematic variable requirements for each signal region are applied to the corresponding control regions.

\begin{figure}[h]
\centering
\includegraphics[width=0.99\linewidth]{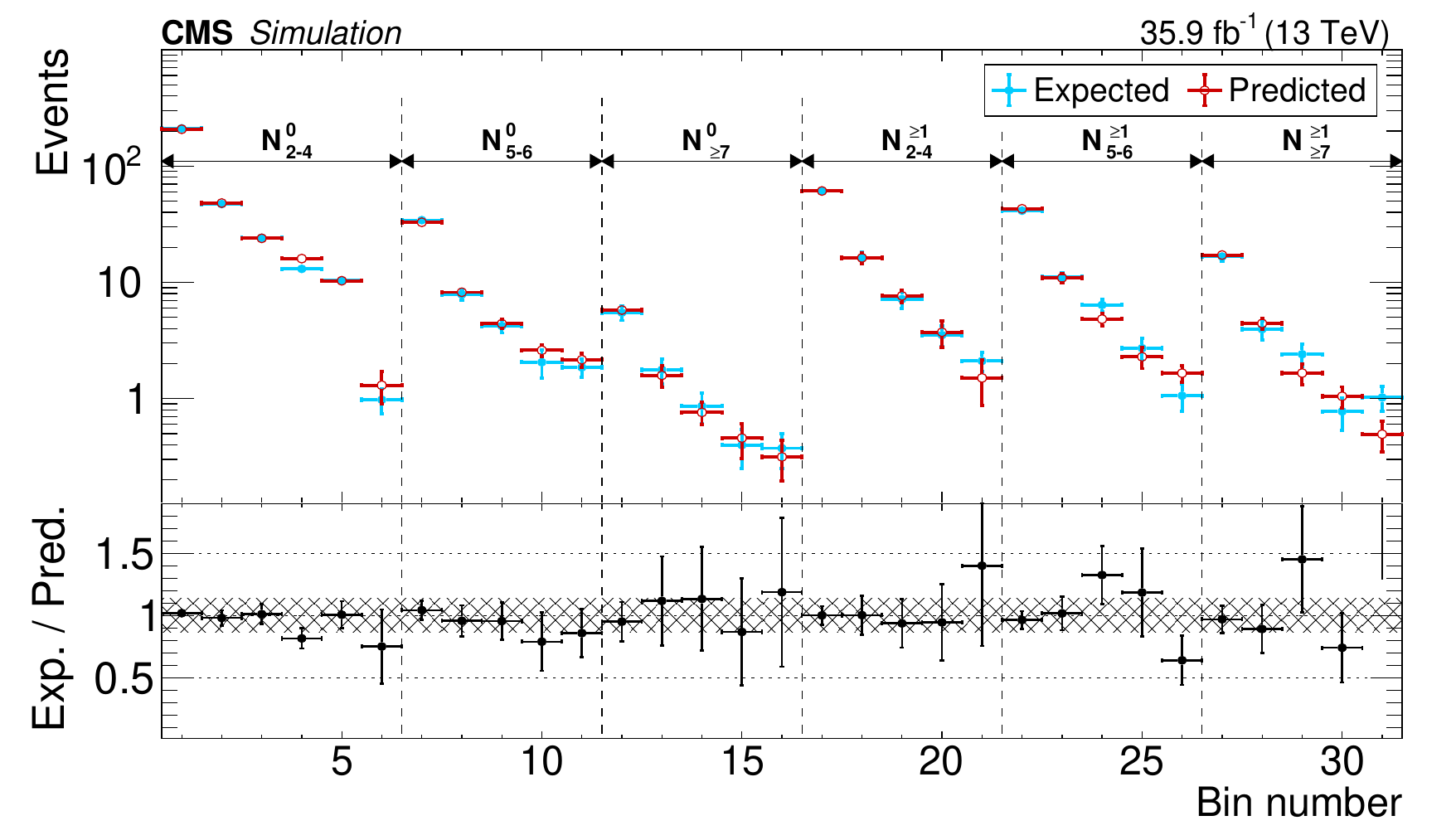}
\caption{The lost-lepton and \tauhad event yields as predicted
directly from simulation in the signal regions, shown in red, and from the prediction procedure applied
to simulated \eleg or \mug events, shown in blue.  The error bars
correspond to the statistical uncertainties from the limited number of events in simulation.  The bottom panel
shows the ratio of the simulation expectation (Exp.) and the simulation-based prediction (Pred.).  The hashed
area shows the expected uncertainties from data-to-simulation correction factors, PDFs, and
renormalization and factorization scales.
The categories, denoted by dashed lines, are labeled as \njb,
where j refers to the number of jets and b refers to the number of {\cPqb}-tagged jets.
The numbered bins within each category are the various \ptmiss bins.
In each of these regions, the first bin corresponds to $100<\ptmiss<200\GeV$, which belongs to a control region.
The remaining bins correspond to the signal regions in Table~\ref{tab:finalPrediction}.}
\label{fig:LostLeptClos}
\end{figure}

Transfer factors are derived using simulated \Wgjets and \ttg processes,
which determine the average number of events
expected in the signal region for each \eleg or \mug event observed in the control region.
The \zg events in which the {\cPZ} boson decays leptonically have a negligible contribution to the transfer factors.
The transfer factors
applied to the \mug control regions account for both lost-{\Pgm} events and
\tauhad events. They are denoted by the symbol \TFmt and are typically in the range ${0.7<\TFmt<1.0}$.  The transfer
factors applied to \eleg events account for only the lost-{\Pe} events.
They are denoted by the symbol \TFe and are typically
in the range ${0.3<\TFe<0.6}$.  The transfer factors are parameterized versus \nj, \nb, and \ptmiss;
however, for ${\ptmiss>150\gev}$, $T_{\ell}$ is found to be independent of \ptmiss.
The parameterization of the transfer factors is validated using simulation by treating \eleg or \mug
events like data and comparing the predicted lost-lepton and \tauhad event yields to the true simulated
event yields in the signal regions.  This comparison is shown in Fig.~\ref{fig:LostLeptClos}.
The prediction in each signal region is ${N_{\ell}^{\text{pred}} = \Sigma_{i} N_{i} T_{\ell,i}}$, where ${\ell=\Pe,\Pgm}$
and $i$ ranges from 1 to $n$, where $n$ is the number of transfer factors that contribute in a given signal region.

\begin{sloppypar} The dominant uncertainty in the lost-lepton background predictions arises from the limited numbers of events in the
\eleg and \mug control regions. These uncertainties are modeled in the final statistical
interpretations as a gamma distribution whose shape parameter is set by the observed
number of events and whose scale parameter is the average transfer factor for that bin. Other
systematic uncertainties in the determination of the transfer factors include the statistical uncertainty
from the limited number of simulated events, which is typically 5--10\% but can be as large as 20\%,
as well as uncertainties in the jet energy corrections, PDFs, renormalization (\muR) and factorization (\muF)
scales, and simulation correction factors. The uncertainties in \muR and \muF are obtained by varying each value independently by factors of 0.5 and 2.0~\cite{Cacciari:2003fi,Catani:2003zt}. Simulation correction factors are used to
account for differences between the observed data and modeling of {\cPqb}-tagging efficiencies,
{\cPqb} jet misidentification, and lepton reconstruction efficiencies in simulation.
One of the largest uncertainties, apart from the statistical uncertainty in the data control regions and the simulation, comes from mismodeling of photons which are collinear with electrons, which has a 12\% effect on the lost-lepton prediction.
\end{sloppypar}

\subsection{Misidentified photon background}

Events containing the decay $\PW\to\Pe\PGn$ are the primary source of electrons that are erroneously identified as photons.
Photon misidentification can occur when a pixel detector seed fails to be associated with the
energy deposit in the ECAL. Given a misidentification rate, which relates events with an erroneously identified photon
to events with a well-identified electron,
the photon background can be estimated from a single-electron (zero-photon) control
region. The misidentification rate is estimated in simulation and corrections are derived from observed data to account
for any mismodeling in simulation.

The single-electron control regions are defined by the same kinematic requirements as the single-photon
signal regions, except that we require no photons and exactly one electron,
and we use the momentum of the electron in place of the momentum of the photon for photon-based variables. As explained in the previous section, in addition to all of the signal region selections, events are required to satisfy ${\mT(\Pe,\ptmiss)<100\gev}$.

To extrapolate from the event yields in the single-electron control regions to the event yields for the misidentified photon background in the
signal regions, we derive a misidentification rate $f=N_{\cPgg}/N_{\Pe}$ using a combination of simulation and data.
The misidentification rate is determined as a function of the electron \pt and the multiplicity \qmult of
charged-particle tracks from the primary vertex in a region around the electron candidate.  The charged-track multiplicity
is computed by counting the number of charged PF candidates (electrons, muons, hadrons) in the jet closest to the electron candidate.
If there is no jet within $\deltar < 0.3$ of the electron candidate, \qmult is set
to zero.  A typical event in the single-electron control region has a \qmult of 3--4.
The electron \pt and \qmult dependence of the misidentification rate is derived using simulated \Wjets and \ttjets events.
The misidentification rate is on average 1--2\%, but can be as low as 0.5\% for events with high \qmult.

\begin{sloppypar} To account for systematic differences between the misidentification rates in data and simulation, we correct the
misidentification rate by measuring it in both simulated and observed Drell--Yan (DY) events.
Separate corrections are derived for low \qmult (${\leq}1$) and high \qmult (${\geq}2$).
The DY control region is defined by requiring one electron with ${\pt>40\GeV}$ and another reconstructed particle, either a photon or an oppositely charged electron,
with ${\pt>100\GeV}$. A further requirement ${50<(m_{\EE}\,\text{or}\,m_{\eleg})<130\GeV}$ is applied to ensure the particles are consistent with the decay products of a {\PZ} boson,
and therefore the photon is likely to be a misidentified electron. The misidentification rate is computed as the ratio $N_{\eleg}/N_{\EE}$,
where $N_{\eleg}$ ($N_{\EE}$) is the number of events in the \eleg (\EE) control region.
It is found to be 15--20\% higher in data than in simulation.
\end{sloppypar}

The prediction of the misidentified-photon background in the signal region is then given by the weighted sum of
the observed events in the control region, where the weight is given by the data-corrected
misidentification rate for photons.  The dominant uncertainty in the prediction is a 14\% uncertainty in the data-to-simulation correction factors, followed by
the uncertainty in the limited number of events in the simulation at large values of \ptmiss.
The misidentified-photon background prediction also includes uncertainties in the modeling of initial-state radiation (ISR)
in the simulation, statistical uncertainties from the limited number of events in the data control
regions, uncertainties in the pileup modeling, and uncertainties in the trigger efficiency
measurement.

\subsection{Background from \texorpdfstring{\znng}{Z(nu nu) + photon} events}

Decays of the {\PZ} boson to invisible particles constitute a major background for events with low \nj, low \nb, and high \ptmiss.
The \znng background is estimated using \zllg events.
The shape of the distribution of \ptmiss vs. \nj in \znng events is modeled in
simulation, while the normalization and the purity of the control region are measured in data.

Events in the \llg control region are required to have exactly two
oppositely charged, same-flavor leptons ($\ell = \Pe$ or \Pgm) and one photon with $\pt>100\gev$.
The dilepton invariant mass $\mdilep$ is required to be consistent with the
\PZ boson mass, $80<\mdilep<100\gev$. The charged leptons
serve as a proxy for neutrinos, so the event-level kinematic variables, such as \ptmiss, are
calculated after removing charged leptons from the event.

The \llg control region may contain a small fraction of events from processes other than \zllg, primarily {\ttbar}\cPgg.
We define the purity of the control region as the percentage of events originating from the \zllg process.
The purity is computed in data by measuring the
number of events in the corresponding oppositely charged, different-flavor control region, which has a higher proportion of \ttg events.
The purity is found to be $(97\pm3)\%$.
A statistically compatible purity is also measured in the oppositely charged, same-flavor
control region. In this region, the $\mdilep$ distribution is used to extrapolate from the number of events with $\mdilep$ far from the {\PZ} boson mass
to the number of events with $\mdilep$ close to it.

\begin{sloppypar} The \znng predictions from simulation are scaled to the total \zllg
yield observed according to ${N_{\znng} = \beta \Rnnll N_{\zllg}}$,
where $\beta$ is the purity of the $\zllg$ control region and $\Rnnll$
is the ratio between the expected number of $\znng$ and $\zllg$ events.
The ratio $\Rnnll$, which accounts for lepton reconstruction effects and
the relative branching fractions for $\cPZ\to\PGn\Pagn$ and $\cPZ\to\ell^{+}\ell^{-}$, is determined from simulation.
\end{sloppypar}

\begin{sloppypar} The primary uncertainty in the $\znng$ prediction arises from uncertainties
in the \ptmiss distribution from the simulation.  Other uncertainties include statistical
uncertainties from the limited number of events in the simulation and uncertainties in the estimation of the control region
purity. The \ptg-dependent NLO electroweak corrections~\cite{Denner:2015fca}
are assigned as additional uncertainties to account for any mismodeling of
the photon \pt in simulation. This uncertainty has a magnitude of 8\% for
the lowest \ptmiss bin and rises to 40\% for $\ptmiss>750\gev$.
\end{sloppypar}

\subsection{Background from \texorpdfstring{\cPgg}{photon}+jets events}

\begin{sloppypar} The \gjets background is dominated by events in which a genuine photon is accompanied by an energetic jet
with mismeasured \pt, resulting in high \ptmiss. The QCD multijet events with a jet misidentified as a photon and
a mismeasured jet contribute to this background at a much smaller rate; these events are
measured together with events from the \gjets process.
Most of these events are removed by requiring that
the azimuthal angles between the \ptvecmiss and each of the two highest \pt jets satisfy $\dphi_{1,2}>0.3$.
Inverting this requirement provides a large control region of low-$\dphi$
events that is used to predict the \gjets background in the signal regions.
The ratio of high-\dphi events to low-\dphi events, \rhl, is derived from
the low-\ptmiss sideband ($100<\ptmiss<200\gev$).
\end{sloppypar}

While most of the events in both the low-\dphi and the low-\ptmiss control regions
are \gjets events, electroweak backgrounds in which \ptmiss arises from {\PW} or \PZ bosons
decaying to one or more neutrinos,
like those discussed previously, will contaminate these control regions.
The contamination can be significant
for high \nj and \nb, where {\ttbar} events are more prevalent.  The rates
of these events in the control regions are predicted using the same techniques, as discussed in the previous sections.

\begin{figure}[h]
\centering
\includegraphics[width=0.99\linewidth]{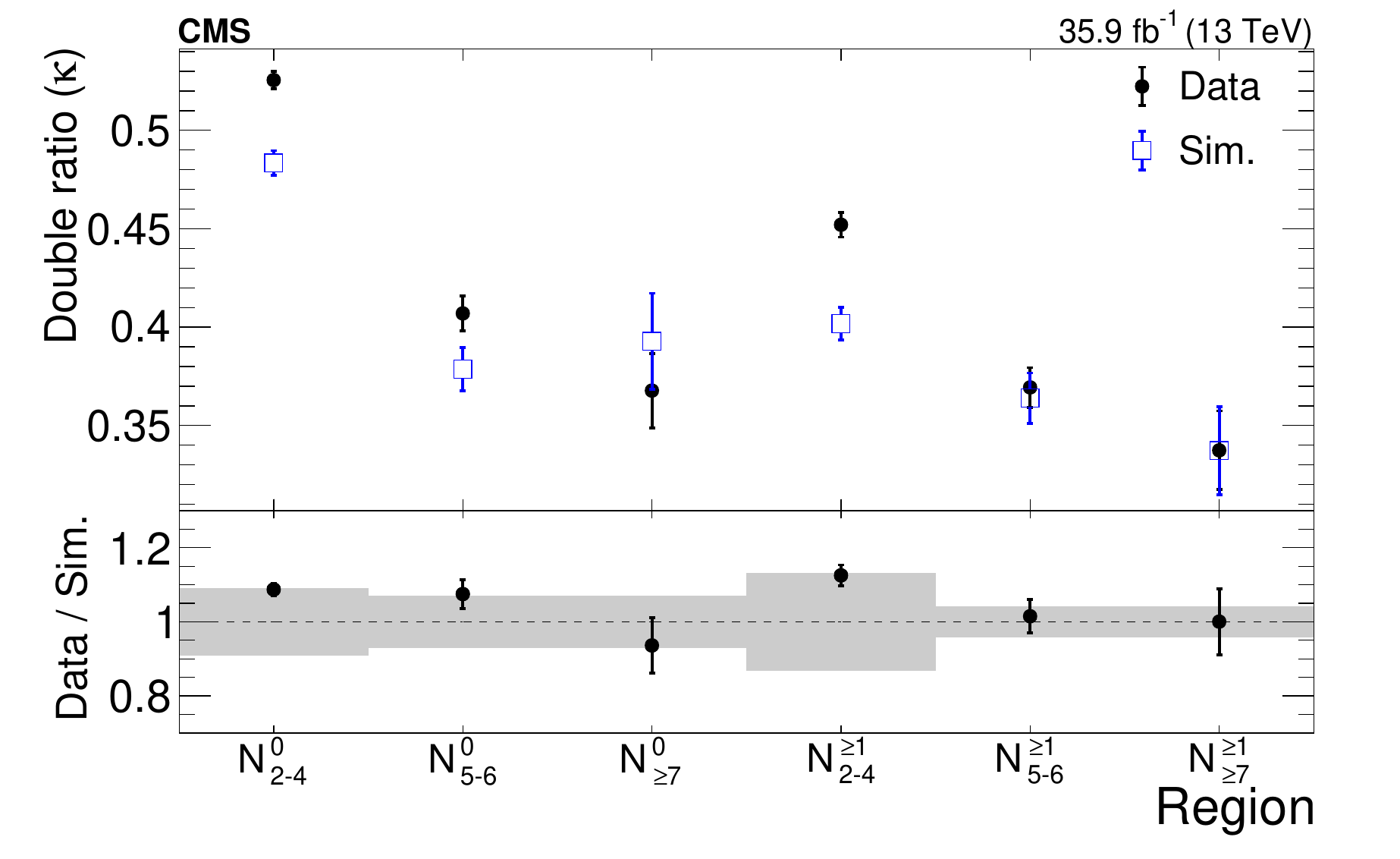}
\caption{The double ratio $\kappa$ in each \nj-\nb region for
zero-photon events. The filled black circles are the observed $\kappa$ values after subtracting
the electroweak contamination based on simulation. The open blue squares are the
$\kappa$ values computed directly from simulation.  The ratio is shown in the bottom
panel, where the shaded region corresponds to the systematic uncertainty in the \gjets prediction.
In the label \njb, j refers to the number of jets and b refers to the number of {\cPqb}-tagged jets.
}
\label{fig:kappa_validation}
\end{figure}

A double ratio $\kappa=\rhl^{\ptmiss>200\GeV}/\rhl^{\ptmiss<200\GeV}$
is derived from simulated \gjets events in order to account for the dependence of \rhl on \ptmiss.
To test how well the simulation models $\kappa$, we use a zero-photon validation region
in which the contribution from events containing a mismeasured jet dominates.
To be consistent with the trigger used to select the data in this region, these events are also required to have $\HT>1000\gev$.
Electroweak contamination in the zero-photon validation region is estimated using simulated
\Vgjets ({\PV} = \cPZ, {\PW}), \ttg, \ttjets, \Wjets, and {\znn}+jets events.
The comparison of $\kappa$ in data and simulation is shown in Fig.~\ref{fig:kappa_validation}.
The level of disagreement is found to be less than 20\%.

Event yields for the \gjets background are computed from the high-\ptmiss,
low-\dphi control regions according to $N_{\gjets} = \kappa N_{\text{low-}\dphi}\rhl$.
$N_{\text{low-}\dphi}$ is the event yield in the high-\ptmiss, low-\dphi control
region after removing contributions from electroweak backgrounds.

Uncertainties in the \gjets prediction are
dominated by the statistical uncertainties either from the limited number of events in the
low-\dphi control regions or from the predictions of the electroweak
contamination. The ${<}20\%$ disagreement between the $\kappa$ values in data and simulation in the zero-photon
validation region is included as an additional uncertainty.
Uncertainties in the {\cPqb}-tagging correction factors are a minor contribution to
the uncertainty in the \gjets prediction.

\section{Results and interpretations}
\label{sec:results}

\begin{figure}[htb!]
\centering
\includegraphics[width=0.99\linewidth]{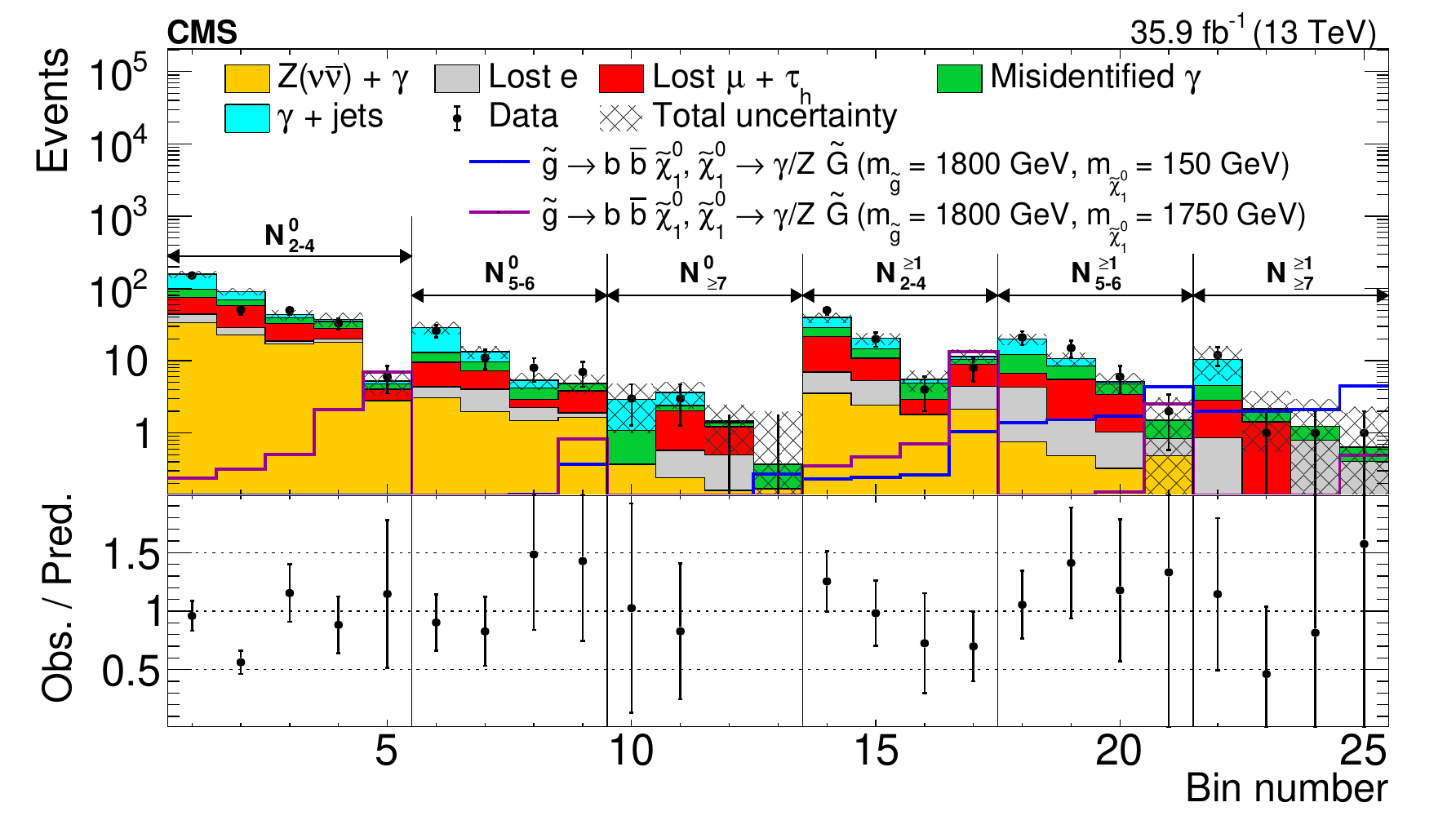}
\caption{Observed numbers of events and predicted numbers of events from the various SM backgrounds in the 25 signal regions.
The categories, denoted by vertical lines, are labeled as \njb,
where j refers to the number of jets and b refers to the number of {\cPqb}-tagged jets.
The numbered bins within each category are the various \ptmiss bins, as defined in Table~\ref{tab:finalPrediction}. The lower panel
shows the ratio of the observed events to the predicted SM background events.
The error bars in the lower panel are the quadrature sum of the statistical uncertainty in the observed data
and the systematic uncertainty in the predicted backgrounds before the adjustments based on
a maximum likelihood fit to data assuming no signal strength.}
\label{fig:summaryPlot}
\end{figure}

The predicted background and observed yields are shown in Table~\ref{tab:finalPrediction} and Fig.~\ref{fig:summaryPlot}.
The largest deviation is found in bin 2 ($2\leq\nj\leq 4$, $\nb=0$, and $270<\ptmiss<350\gev$), where the background is predicted to be 91 events
with 51 events observed.
The local significance of this single bin was computed to be around 2 standard deviations below the SM expectation.
This calculation does not account for the look-elsewhere effect associated with the use of 25 exclusive signal regions,
which is expected to reduce this significance.
In general, a large deviation in a single bin is inconsistent with the expected distributions of events from the signal models considered here.
The observations in all other bins are consistent with the SM expectations within one standard deviation.

\begin{table*}[htb!]
\renewcommand{\arraystretch}{1.25}
\centering
\topcaption{Predicted and observed event yields for each of the 25 exclusive signal regions.}
\label{tab:finalPrediction}
\cmsTable{\begin{tabular}{cccccccccc}
\nj      & \nb& \ptmiss [GeV]& Lost \Pe              & Lost {\Pgm} + \tauhad & Misid. \cPgg   & \znng                & \gjets            & Total                 & Data \\
\hline
2--4     & 0  & 200--270     & 10.5 $\pm$ 2.6        & 31.2 $\pm$ 6.0        & 22.3 $\pm$ 5.4 & 33.6 $\pm$ 8.3       & 60   $\pm$ 11         & 157 $\pm$ 16          & 151 \\
2--4     & 0  & 270--350     & 5.8  $\pm$ 1.8        & 29.6 $\pm$ 5.9        & 11.9 $\pm$ 2.9 & 22.9 $\pm$ 6.0       & 20.5 $\pm$ 4.3        & 91  $\pm$ 10          & 51 \\
2--4     & 0  & 350--450     & 1.68 $\pm$ 0.88       & 13.9 $\pm$ 3.9        & 6.6 $\pm$ 1.6  & 17.0 $\pm$ 5.2       & 4.1  $\pm$ 1.4        & 43.3$\pm$ 6.8         & 50 \\
2--4     & 0  & 450--750     & 1.98 $\pm$ 0.94       & 8.1  $\pm$ 3.1        & 6.7 $\pm$ 1.5  & 18.1 $\pm$ 7.1       & 2.5  $\pm$ 1.3        & 37.4$\pm$ 8.0         & 33 \\
2--4     & 0  & ${>}750$     & $0.00_{-0.00}^{+0.69}$& 1.2 $\pm$ 1.2         & 0.79 $\pm$ 0.19& 2.8 $\pm$ 1.2        & $0.41_{-0.41}^{+0.42}$& 5.2 $\pm$ 1.9         & 6 \\
\\[\cmsTabSkip]
5--6     & 0  & 200--270     & 1.28 $\pm$ 0.61       & 5.1  $\pm$ 1.9        & 3.53 $\pm$ 0.75& 3.09 $\pm$ 0.78      & 15.8 $\pm$ 4.8        & 28.8 $\pm$ 5.3        & 26 \\
5--6     & 0  & 270--350     & 2.06 $\pm$ 0.80       & 3.2  $\pm$ 1.5        & 2.39 $\pm$ 0.56& 1.98 $\pm$ 0.54      & 3.7  $\pm$ 1.8        & 13.3 $\pm$ 2.6        & 11  \\
5--6     & 0  & 350--450     & 0.77 $\pm$ 0.46       & $0.64_{-0.64}^{+0.65}$& 1.26 $\pm$ 0.30& 1.49 $\pm$ 0.47      & 1.23 $\pm$ 0.97       & 5.4 $\pm$ 1.4         & 8 \\
5--6     & 0  & ${>}450$     & 0.26 $\pm$ 0.26       & 1.9 $\pm$ 1.1         & 1.00 $\pm$ 0.24& 1.65 $\pm$ 0.65      & $0.07_{-0.07}^{+0.52}$& 4.9 $\pm$ 1.4         & 7 \\
\\[\cmsTabSkip]
${\geq}7$& 0  & 200--270     & $0.00_{-0.00}^{+0.61}$& $0.0_{-0.0}^{+1.3}$   & 0.72 $\pm$ 0.16& 0.37 $\pm$ 0.11      & 1.8 $\pm$ 1.2         & 2.9 $\pm$ 1.9         & 3 \\
${\geq}7$& 0  & 270--350     & $0.34_{-0.34}^{+0.35}$& 1.5 $\pm$ 1.0         & 0.38 $\pm$ 0.10& 0.24 $\pm$ 0.08      & 1.22 $\pm$ 0.94       & 3.6 $\pm$ 1.5         & 3 \\
${\geq}7$& 0  & 350--450     & $0.34_{-0.34}^{+0.35}$& 0.73 $\pm$ 0.73       & 0.17 $\pm$ 0.05& 0.16 $\pm$ 0.07      & $0.07_{-0.07}^{+0.50}$& 1.46 $\pm$ 0.96       & 0 \\
${\geq}7$& 0  & ${>}450$     & $0.00_{-0.00}^{+0.61}$& $0.0_{-0.0}^{+1.3}$   & 0.20 $\pm$ 0.06& 0.17 $\pm$ 0.08      & $0.00_{-0.00}^{+0.75}$& $0.37_{-0.37}^{+1.60}$& 0 \\
\\[\cmsTabSkip]
2--4     & ${\geq}1$  & 200--270     & 3.4  $\pm$ 1.5        & 14.5 $\pm$ 4.2        & 7.1  $\pm$ 1.7 & 3.55 $\pm$ 0.89      & 11.3 $\pm$ 3.3        & 39.8  $\pm$ 5.9       & 50 \\
2--4     & ${\geq}1$  & 270--350     & 2.9  $\pm$ 1.4        & 5.6  $\pm$ 2.5        & 3.79 $\pm$ 0.92& 2.45 $\pm$ 0.65      & 5.7  $\pm$ 1.8        & 20.4  $\pm$ 3.6       & 20   \\
2--4     & ${\geq}1$  & 350--450     & $0.0_{-0.0}^{+1.0}$   & 1.1 $\pm$ 1.1         & 2.00 $\pm$ 0.45& 1.81 $\pm$ 0.55      & 0.59 $\pm$ 0.44       & 5.5  $\pm$ 1.7        & 4  \\
2--4     & ${\geq}1$  & ${>}450$     & 2.3 $\pm$ 1.2         & 4.4 $\pm$ 2.3         & 1.62 $\pm$ 0.38& 2.14 $\pm$ 0.84      & 0.95 $\pm$ 0.54       & 11.5 $\pm$ 2.8        & 8  \\
\\[\cmsTabSkip]
5--6     & ${\geq}1$  & 200--270     & 3.5  $\pm$ 1.3        & 2.4  $\pm$ 1.4        & 5.5  $\pm$ 1.2 & 0.76 $\pm$ 0.20      & 7.7  $\pm$ 2.4        & 19.9  $\pm$ 3.3       & 21 \\
5--6     & ${\geq}1$  & 270--350     & 1.06 $\pm$ 0.64       & 4.0  $\pm$ 1.8        & 2.98 $\pm$ 0.63& 0.49 $\pm$ 0.14      & 2.1 $\pm$ 1.0         & 10.6  $\pm$ 2.3       & 15 \\
5--6     & ${\geq}1$  & 350--450     & 0.71 $\pm$ 0.51       & 2.4  $\pm$ 1.4        & 1.38 $\pm$ 0.29& 0.32 $\pm$ 0.11      & $0.30_{-0.30}^{+0.49}$& 5.1 $\pm$ 1.6         & 6   \\
5--6     & ${\geq}1$  & ${>}450$     & $0.35_{-0.35}^{+0.36}$& $0.0_{-0.0}^{+1.4}$   & 0.67 $\pm$ 0.15& 0.48 $\pm$ 0.20      & $0.00_{-0.00}^{+0.56}$& $1.5_{-1.5}^{+1.6}$   & 2 \\
\\[\cmsTabSkip]
${\geq}7$& ${\geq}1$  & 200--270     & 0.72 $\pm$ 0.53       & 2.0 $\pm$ 1.2         & 1.68 $\pm$ 0.37& 0.13 $\pm$ 0.04      & 5.9 $\pm$ 5.0         & 10.5 $\pm$ 5.1        & 12 \\
${\geq}7$& ${\geq}1$  & 270--350     & $0.00_{-0.00}^{+0.65}$& 1.33 $\pm$ 0.96       & 0.73 $\pm$ 0.16& 0.10 $\pm$ 0.04      & $0.0_{-0.0}^{+1.1}$   & 2.2 $\pm$ 1.6         & 1 \\
${\geq}7$& ${\geq}1$  & 350--450     & 0.72 $\pm$ 0.53       & $0.0_{-0.0}^{+1.2}$   & 0.44 $\pm$ 0.10& 0.07 $\pm$ 0.03      & $0.0_{-0.0}^{+1.1}$   & $1.2_{-1.2}^{+1.7}$   & 1 \\
${\geq}7$& ${\geq}1$  & ${>}450$     & $0.36_{-0.36}^{+0.37}$& $0.0_{-0.0}^{+1.2}$   & 0.23 $\pm$ 0.07& 0.04 $\pm$ 0.02      & $0.0_{-0.0}^{+1.1}$   & $0.6_{-0.6}^{+1.7}$   & 1 \\
\end{tabular}}
\end{table*}

\begin{sloppypar} Limits are evaluated for the production cross sections of the signal scenarios discussed in Section~\ref{sec:introduction}
using a maximum likelihood fit for the SUSY signal strength, the yields of the five classes of background events shown
in Fig.~\ref{fig:summaryPlot}, and various nuisance parameters.
The SUSY signal strength $\mu$ is defined to be the ratio of the observed signal cross section to the predicted cross section.
A nuisance parameter refers to a variable not of interest in this search, such as the effect of parton distribution function
uncertainties in a background prediction. The nuisance parameters are constrained by observed data in the fit.
The uncertainties in the predicted signal yield arise from the uncertainties in renormalization and factorization scales, ISR modeling,
jet energy scale, {\cPqb}-tagging efficiency and misidentification rate, corrections to simulation, limited numbers of simulated events, and the integrated luminosity measurement~\cite{CMS-PAS-LUM-17-001}. The largest uncertainty
comes from the ISR modeling; it ranges from 4 to 30\% depending on the signal region and the signal parameters, taking higher values
for regions with large \nj or for signals with $\deltam \approx 0$. Here, \deltam is the difference in mass between the gluino or squark and its decay products,
e.g. $\deltam = \mgluino - (\mnlsp + 2m_{\cPqt})$ for the T5ttttZG model when on-shell top quarks are produced.
The second-largest uncertainty comes from the correction for differences between \GEANTfour
and the fast simulation in \ptmiss modeling, with a maximum value of 10\%. The procedures used to evaluate the systematic uncertainties
in the signal predictions in the context of this search are described in Ref.~\cite{Sirunyan:2017cwe}.
\end{sloppypar}

For the models of gluino pair production considered here, the limits are derived as a function of \mgluino and \mnlsp,
while for the model of top squark pair production, the limits are a function of \mstop and \mnlsp.
The likelihood used for the statistical interpretation models the yield in each of the signal regions as
a Poisson distribution, multiplied by constraints which account for the uncertainties in the
background predictions and signal yields.  For the predictions in which an observed event yield
in a control region is scaled, a gamma distribution is used to model the Poisson
uncertainty of the observed control region yield.  All other uncertainties are modeled as log-normal
distributions. The test statistic is $q_{\mu}=-2\ln{\mathcal{L}_{\mu}/\mathcal{L}_{\text{max}}}$, where $\mathcal{L}_{\text{max}}$
is the maximum likelihood determined by leaving all parameters as free, including the signal strength,
and $\mathcal{L}_{\mu}$ is the maximum likelihood for a fixed value of $\mu$. Limits are determined using an
approximation of the asymptotic form of the test statistic distribution~\cite{Cowan:2010js} in conjunction with the \CLs
criterion~\cite{Junk1999,Read:2002hq}.  Expected upper limits are derived by varying observed yields according
to the expectations from the background-only hypothesis.

\begin{figure*}[htb!]
\centering
\includegraphics[width=0.48\linewidth]{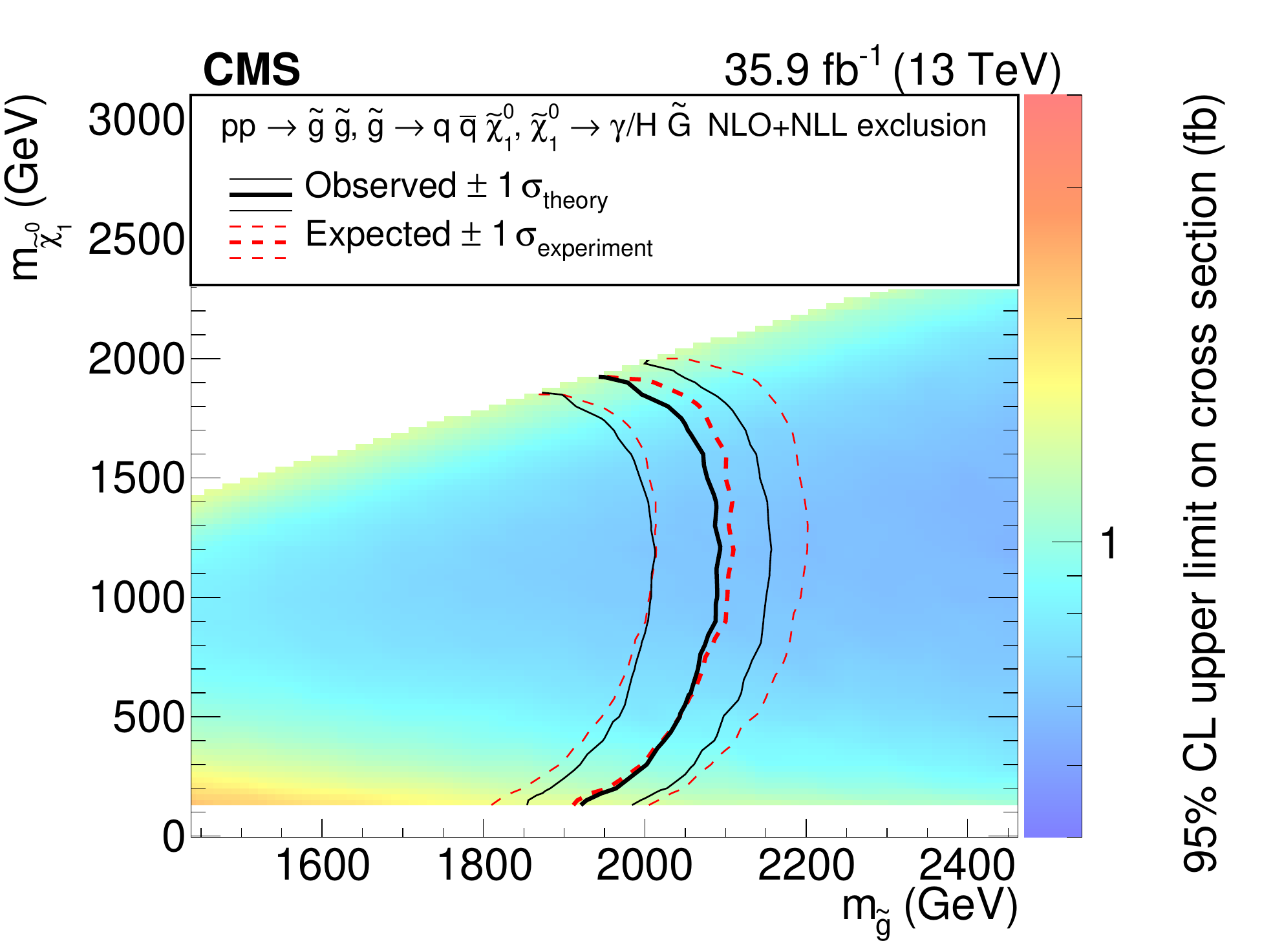}
\includegraphics[width=0.48\linewidth]{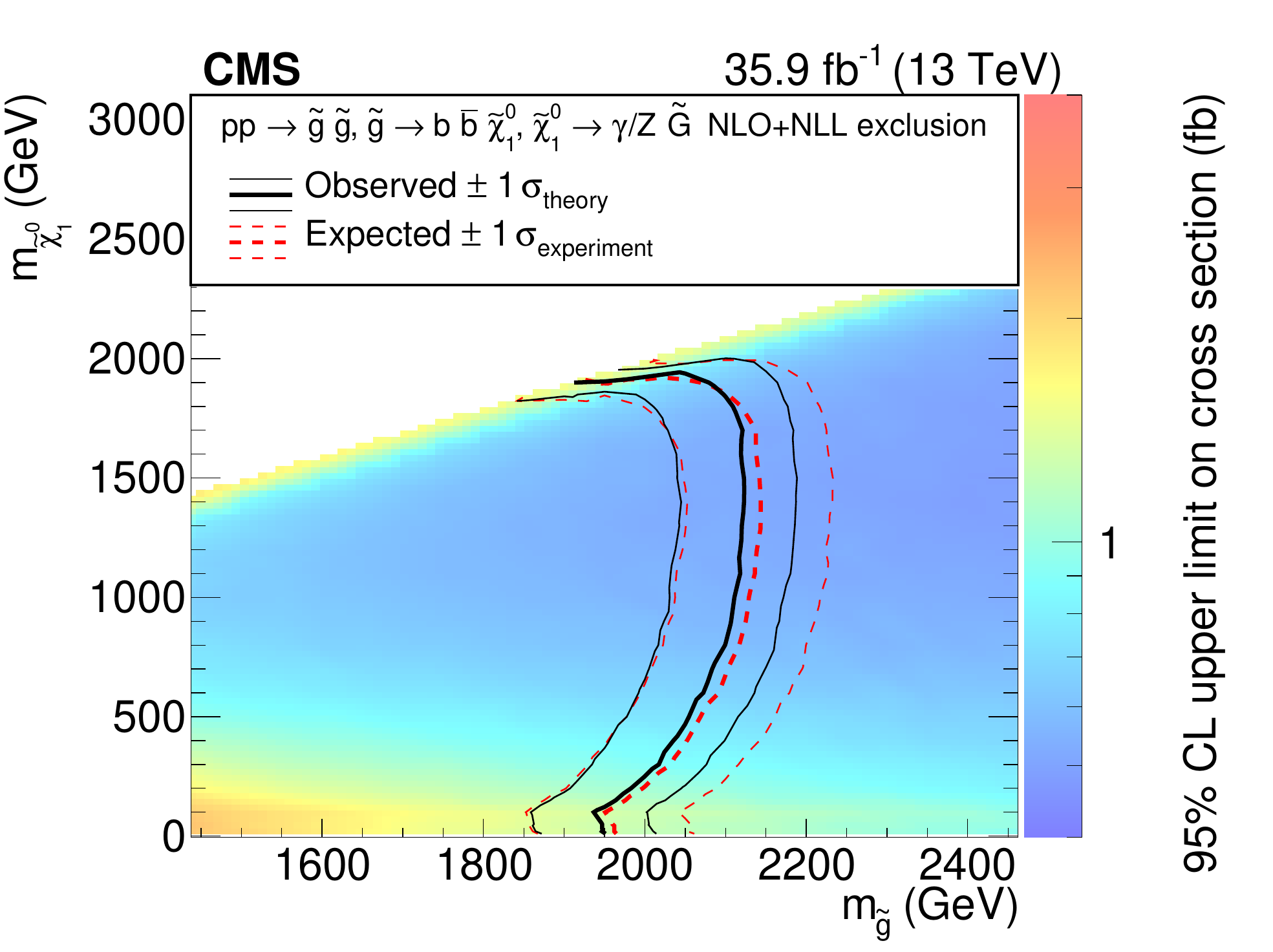}\\
\includegraphics[width=0.48\linewidth]{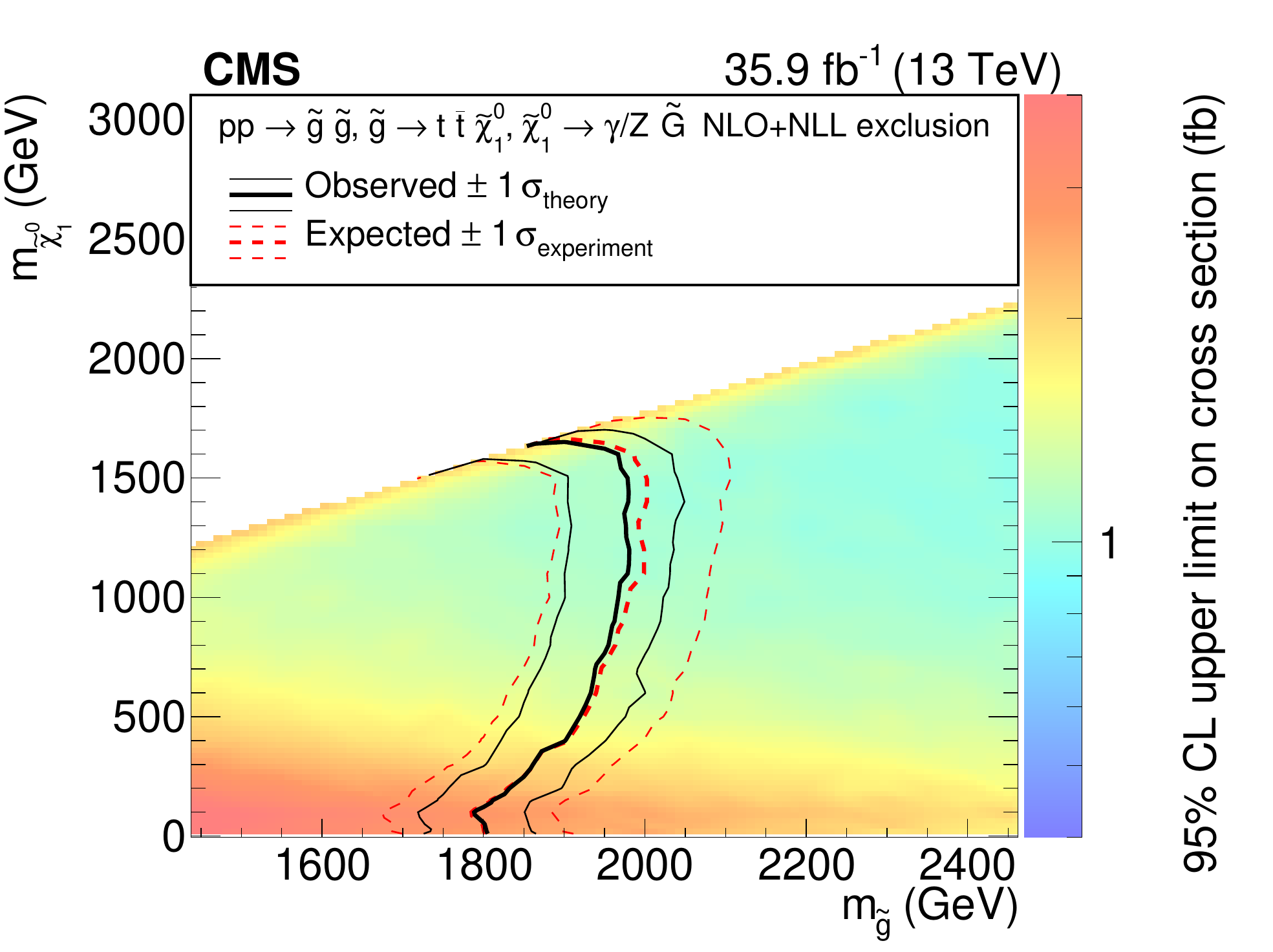}
\includegraphics[width=0.48\linewidth]{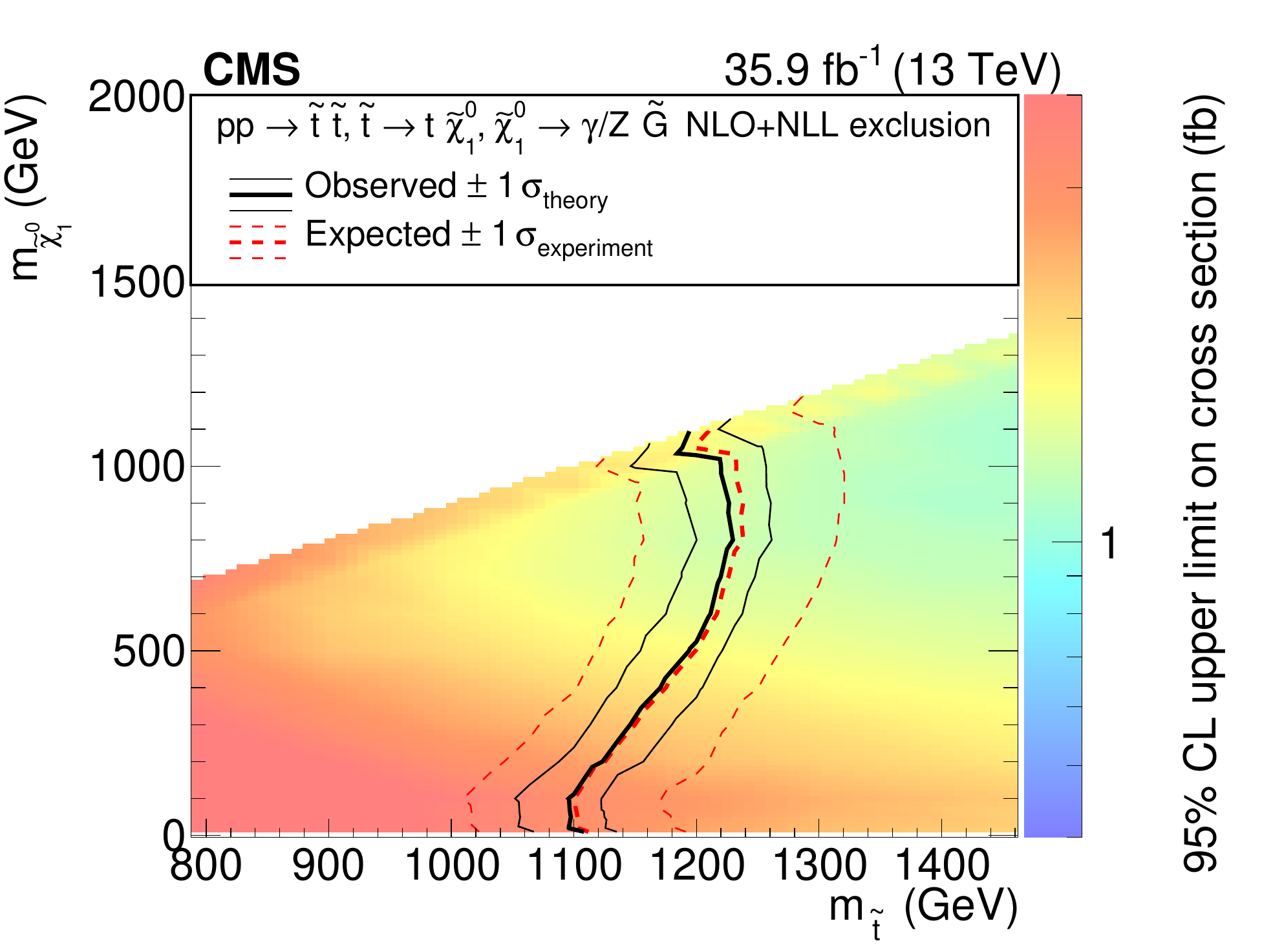}
\caption{Observed and expected 95\% \CL upper limits for gluino or top squark pair production cross sections
for the T5qqqqHG (upper left), T5bbbbZG (upper right), T5ttttZG (bottom left), and T6ttZG
(bottom right) models. Black lines denote the
observed exclusion limit and the uncertainty due to variations of the
theoretical prediction of the gluino or top squark pair production cross section.
The dashed lines correspond to the region containing 68\% of the distribution of
the expected exclusion limits under the background-only hypothesis.}
\label{fig:exclusions}
\end{figure*}

\begin{sloppypar} Using the statistical procedure described above, 95\% confidence level (\CL) upper limits
are computed on the signal cross section for each simplified model and each mass hypothesis.
Exclusion limits are defined by comparing observed upper limits to the predicted
NLO+NLL signal cross section. The signal cross
sections are also varied according to theoretical uncertainties to give a ${\pm}1$ standard deviation variation
on the observed exclusion contour.  The 95\% \CL cross section limits and exclusion contours for the four models considered, T5qqqqHG, T5bbbbZG,
T5ttttZG, and T6ttZG, are shown in Fig.~\ref{fig:exclusions}.
\end{sloppypar}

Generally, the limits degrade at both high and low $\mnlsp$.
For $\mnlsp\approx \mgluino\,(\mstop)$, the quarks from the decay of gluinos (top squarks) have low \pt.
Correspondingly, the \htg, \nj, and \nb distributions tend toward lower values, reducing the signal efficiency and causing
signal events to populate regions with higher background yields. For small $\mnlsp$,
the quarks produced in the decay of gluinos or top squarks have high \pt but lower \ptmiss on average.
For all models except T5qqqqHG, when the NLSP mass drops below the mass of the
{\PZ} boson, the kinematics of the NLSP decay require the {\PZ} boson to be far off-shell.
As the {\PZ} boson mass is forced to be lower, the LSP will carry a larger fraction of the momentum of the NLSP, producing larger \ptmiss.
This causes a slight increase in the sensitivity when the NLSP mass is near the {\PZ} boson mass.
While a similar effect would happen for the T5qqqqHG model,
the simulation used here does not probe the region of parameter space where the Higgs boson
would be forced to have a mass far off-shell.
Similarly, the limits for top squark production improve slightly at very high $\mnlsp$, when the top quarks become off-shell.
In this case, the \PSGczDo carries a larger fraction of the top squark momentum, increasing the \ptmiss.

\begin{sloppypar} For moderate \mnlsp, gluino masses as large as 2090, 2120, and 1970\GeV are excluded for the T5qqqqHG, T5bbbbZG, and T5ttttZG models, respectively.
Top squark masses as large as 1230\GeV are excluded for the T6ttZG model.
For small \mnlsp, gluino masses as large as 1920, 1950, and 1800\GeV are excluded for the T5qqqqHG, T5bbbbZG, and T5ttttZG models, respectively.
Top squark masses as large as 1110\GeV are excluded for the T6ttZG model.
There is close agreement between the observed and expected limits.
\end{sloppypar}

\section{Summary}
\label{sec:summary}

\begin{sloppypar} A search for gluino and top squark pair production is presented, based on a proton-proton collision
dataset at a center-of-mass energy of 13\TeV recorded with the CMS detector in 2016.
The data correspond to an integrated luminosity of 35.9\fbinv.  Events are
required to have at least one
isolated photon with transverse momentum ${\pt>100\gev}$, two jets with ${\pt>30\gev}$ and pseudorapidity ${\abseta<2.4}$,
and missing transverse momentum ${\ptmiss>200\gev}$.
\end{sloppypar}

The data are categorized into 25 exclusive signal regions based on the number of
jets, the number of {\cPqb}-tagged jets, and \ptmiss.  Background yields from the standard model
processes are predicted using simulation and data control regions.  The observed event
yields are found to be consistent with expectations from the standard model processes
within the uncertainties.

Results are interpreted in the context of simplified models.  Four such models
are studied, three of which involve gluino pair production
and one of which involves top squark pair production. All models assume
a gauge-mediated supersymmetry (SUSY) breaking scenario, in which the lightest SUSY particle is a gravitino (\PXXSG).
We consider scenarios in which the gluino decays to
a neutralino \PSGczDo and a pair of light-flavor quarks (T5qqqqHG),
bottom quarks (T5bbbbZG), or top quarks (T5ttttZG).
In the T5qqqqHG model, the \PSGczDo decays with equal probability either to a photon and a \PXXSG
or to a Higgs boson and a \PXXSG.  In the T5bbbbZG
and T5ttttZG models, the \PSGczDo decays with equal probability either to a photon and a \PXXSG or to a
\PZ boson and a \PXXSG.  In the top squark pair production model (T6ttZG),
top squarks decay to a top quark and \PSGczDo, and the \PSGczDo decays with equal probability either to a photon and a \PXXSG or to a \PZ boson and a \PXXSG.

Using the cross sections for SUSY pair production
calculated at next-to-leading order plus next-to-leading logarithmic accuracy, we place 95\% confidence level lower limits on
the gluino mass as large as 2120\gev, depending on the model and the \mnlsp value, and
limits on the top squark mass as large as 1230\gev, depending on the \mnlsp value.
These results significantly improve upon those from previous searches for SUSY with photons.

\begin{acknowledgments}
We congratulate our colleagues in the CERN accelerator departments for the excellent performance of the LHC and thank the technical and administrative staffs at CERN and at other CMS institutes for their contributions to the success of the CMS effort. In addition, we gratefully acknowledge the computing centers and personnel of the Worldwide LHC Computing Grid for delivering so effectively the computing infrastructure essential to our analyses. Finally, we acknowledge the enduring support for the construction and operation of the LHC and the CMS detector provided by the following funding agencies: BMBWF and FWF (Austria); FNRS and FWO (Belgium); CNPq, CAPES, FAPERJ, FAPERGS, and FAPESP (Brazil); MES (Bulgaria); CERN; CAS, MoST, and NSFC (China); COLCIENCIAS (Colombia); MSES and CSF (Croatia); RPF (Cyprus); SENESCYT (Ecuador); MoER, ERC IUT, and ERDF (Estonia); Academy of Finland, MEC, and HIP (Finland); CEA and CNRS/IN2P3 (France); BMBF, DFG, and HGF (Germany); GSRT (Greece); NKFIA (Hungary); DAE and DST (India); IPM (Iran); SFI (Ireland); INFN (Italy); MSIP and NRF (Republic of Korea); MES (Latvia); LAS (Lithuania); MOE and UM (Malaysia); BUAP, CINVESTAV, CONACYT, LNS, SEP, and UASLP-FAI (Mexico); MOS (Montenegro); MBIE (New Zealand); PAEC (Pakistan); MSHE and NSC (Poland); FCT (Portugal); JINR (Dubna); MON, RosAtom, RAS, RFBR, and NRC KI (Russia); MESTD (Serbia); SEIDI, CPAN, PCTI, and FEDER (Spain); MOSTR (Sri Lanka); Swiss Funding Agencies (Switzerland); MST (Taipei); ThEPCenter, IPST, STAR, and NSTDA (Thailand); TUBITAK and TAEK (Turkey); NASU and SFFR (Ukraine); STFC (United Kingdom); DOE and NSF (USA).

\hyphenation{Rachada-pisek} Individuals have received support from the Marie-Curie program and the European Research Council and Horizon 2020 Grant, contract No. 675440 (European Union); the Leventis Foundation; the A. P. Sloan Foundation; the Alexander von Humboldt Foundation; the Belgian Federal Science Policy Office; the Fonds pour la Formation \`a la Recherche dans l'Industrie et dans l'Agriculture (FRIA-Belgium); the Agentschap voor Innovatie door Wetenschap en Technologie (IWT-Belgium); the F.R.S.-FNRS and FWO (Belgium) under the ``Excellence of Science - EOS" - be.h project n. 30820817; the Ministry of Education, Youth and Sports (MEYS) of the Czech Republic; the Lend\"ulet (``Momentum") Programme and the J\'anos Bolyai Research Scholarship of the Hungarian Academy of Sciences, the New National Excellence Program \'UNKP, the NKFIA research grants 123842, 123959, 124845, 124850 and 125105 (Hungary); the Council of Science and Industrial Research, India; the HOMING PLUS program of the Foundation for Polish Science, cofinanced from European Union, Regional Development Fund, the Mobility Plus program of the Ministry of Science and Higher Education, the National Science Center (Poland), contracts Harmonia 2014/14/M/ST2/00428, Opus 2014/13/B/ST2/02543, 2014/15/B/ST2/03998, and 2015/19/B/ST2/02861, Sonata-bis 2012/07/E/ST2/01406; the National Priorities Research Program by Qatar National Research Fund; the Programa Estatal de Fomento de la Investigaci{\'o}n Cient{\'i}fica y T{\'e}cnica de Excelencia Mar\'{\i}a de Maeztu, grant MDM-2015-0509 and the Programa Severo Ochoa del Principado de Asturias; the Thalis and Aristeia programs cofinanced by EU-ESF and the Greek NSRF; the Rachadapisek Sompot Fund for Postdoctoral Fellowship, Chulalongkorn University and the Chulalongkorn Academic into Its 2nd Century Project Advancement Project (Thailand); the Welch Foundation, contract C-1845; and the Weston Havens Foundation (USA).
\end{acknowledgments}

\clearpage
\bibliography{auto_generated}

\cleardoublepage \appendix\section{The CMS Collaboration \label{app:collab}}\begin{sloppypar}\hyphenpenalty=5000\widowpenalty=500\clubpenalty=5000\vskip\cmsinstskip
\textbf{Yerevan Physics Institute, Yerevan, Armenia}\\*[0pt]
A.M.~Sirunyan, A.~Tumasyan
\vskip\cmsinstskip
\textbf{Institut f\"{u}r Hochenergiephysik, Wien, Austria}\\*[0pt]
W.~Adam, F.~Ambrogi, E.~Asilar, T.~Bergauer, J.~Brandstetter, M.~Dragicevic, J.~Er\"{o}, A.~Escalante~Del~Valle, M.~Flechl, R.~Fr\"{u}hwirth\cmsAuthorMark{1}, V.M.~Ghete, J.~Hrubec, M.~Jeitler\cmsAuthorMark{1}, N.~Krammer, I.~Kr\"{a}tschmer, D.~Liko, T.~Madlener, I.~Mikulec, N.~Rad, H.~Rohringer, J.~Schieck\cmsAuthorMark{1}, R.~Sch\"{o}fbeck, M.~Spanring, D.~Spitzbart, W.~Waltenberger, J.~Wittmann, C.-E.~Wulz\cmsAuthorMark{1}, M.~Zarucki
% ...
\vskip\cmsinstskip
% ...
1:  Also at Vienna University of Technology, Vienna, Austria\\
% ...
\end{sloppypar}
\end{document}